\documentclass{article}

\usepackage{PRIMEarxiv}

\errorcontextlines\maxdimen

\usepackage[utf8]{inputenc} 
\usepackage[T1]{fontenc}    
\usepackage{url}            
\usepackage{booktabs}       
\usepackage{nicefrac}       
\usepackage{microtype}      
\usepackage{fancyhdr}       
\usepackage{graphicx}
\usepackage{xcolor}
\usepackage{threeparttable}
\usepackage{booktabs}
\usepackage{multirow}
\usepackage{tabularx}
\usepackage{arydshln}

\usepackage{lineno,hyperref}
\hypersetup{colorlinks, citecolor=blue, linkcolor=red, urlcolor=green}
\usepackage{amsmath,amsfonts,amssymb}
\usepackage{graphics}
\usepackage{bm}
\usepackage{multirow}

\usepackage{mathtools}
\usepackage{siunitx}
\DeclarePairedDelimiterXPP\BigOSI[2]%
  {\mathcal{O}}{(}{)}{}%
  {\SI{#1}{#2}}

\usepackage{adjustbox}
\usepackage{multirow}
\usepackage{graphicx}
\usepackage{bm}
\usepackage{array}
\usepackage{caption}
\usepackage{subcaption}
\usepackage[ruled, lined, linesnumbered, commentsnumbered, longend]{algorithm2e}
\usepackage{soul}
\usepackage[numbers,sort&compress]{natbib}

\SetKwComment{Comment}{$\triangleright$\ }{}

\title{\underline{\textbf{Fu}}sion-based \underline{\textbf{C}}onstitutiv\underline{\textbf{e}} model (FuCe): Towards model-data augmentation in constitutive modelling}

\author{Tushar \\
  Department of Applied Mechanics\\
  Indian Institute of Technology Delhi\\
  \texttt{amz218314.iitd@gmail.com} \\
  \And
  Sawan Kumar \\
  Department of Applied Mechanics\\
  Indian Institute of Technology Delhi\\
  \texttt{sawan.kumar@am.iitd.ac.in} \\
  \And
 \And
  Souvik Chakraborty \\
  Department of Applied Mechanics\\
  Yardi School of Artificial Intelligence (ScAI)\\
  Indian Institute of Technology Delhi\\
  \texttt{souvik@am.iitd.ac.in} \\
}
\begin{document}
\maketitle
\begin{abstract}
Constitutive modelling is crucial for engineering design and simulations to accurately describe material behavior. However, traditional phenomenological models often struggle to capture the complexities of real materials under varying stress conditions due to their fixed forms and limited parameters. While recent advances in deep learning have addressed some limitations of classical models, purely data-driven methods tend to require large datasets, lack interpretability, and struggle to generalize beyond their training data.
To tackle these issues, we introduce "Fusion-based Constitutive model (FuCe): Towards model-data augmentation in constitutive modelling". This approach combines established phenomenological models with an ICNN architecture, designed to train on the limited and noisy force-displacement data typically available in practical applications. The hybrid model inherently adheres to necessary constitutive conditions. During inference, Monte Carlo dropout is employed to generate Bayesian predictions, providing mean values and confidence intervals that quantify uncertainty.
We demonstrate the model's effectiveness by learning two isotropic constitutive models and one anisotropic model with a single fibre direction, across six different stress states. The framework's applicability is also showcased in finite element simulations across three geometries of varying complexities. Our results highlight the framework's superior extrapolation capabilities, even when trained on limited and noisy data, delivering accurate and physically meaningful predictions across all numerical examples.


\end{abstract}
\section{Introduction}
In simulations of physical systems, constitutive modelling is essential for characterizing the mechanical behaviour of materials. The accuracy of these simulations heavily depends on choosing the right constitutive model \cite{fern2016role}. While some models are grounded in physics, most are primarily phenomenological.
These models work well for small deformations in materials with a linear behaviour. However, when dealing with large deformations in materials like rubbers, polymers, foams, and biological tissues -- which exhibit non-linear and complex behaviours, developing accurate empirical or phenomenological models becomes significantly more difficult. Even with careful formulation and calibration, these models are limited by their fixed forms and the small number of parameters, restricting their ability to accurately represent a wide range of deformation levels and stress states.
Machine learning-based data-driven approach \cite{kirchdoerfer2016data} including methods like Gaussian process regression \cite{williams2006gaussian, fuhg2022local} or neural networks (NNs) \cite{bishop2023deep, kollmannsberger2021deep} presents reasonable alternatives to overcome the time-consuming task of formulating classical constitutive models and to improve the limited expressibility associated with these phenomenological models. The concept of using NNs in constitutive modelling was first presented in the early 1990s \cite{ghaboussi1991knowledge}. The fundamental concept of machine-learning-based material modeling involves choosing a highly expressive model approach, characterized by a large number of learnable parameters, and then tuning these parameters using the available data. However, in this early phase, black-box approaches were mostly used, i.e., networks that do not take into account any physical principles and therefore
only perform satisfactorily in the training regime. However, these models offer poor generalization beyond the training data regime.
{In addition, a large dataset of stress-strain pairs is required to train the constitutive models. This presents two new issues when using experimental data. First, it is impractical to investigate the complete high-dimensional stress-strain space via conventional mechanical experiments, such as uniaxial or biaxial tensile and bending tests. Second, constructing comprehensive constitutive tensorial models is challenging since stress tensors are not directly measurable, and force measurements only represent boundary average projections of stress tensors, as highlighted in \cite{joshi2022bayesian}. Although multiscale simulations can produce large, detailed datasets of stress-strain relationships \cite{yuan2008toward, wang2018multiscale}, their computational demands for complex material systems remain prohibitive. Therefore, it is crucial to learn material behaviour directly from the force-displacement data that is practically available through mechanical testing.}

{There have been attempts towards discovering constitutive relations from force-displacement data. For example, EUCLID and its variants \cite{thakolkaran2022nn,joshi2022bayesian,flaschel2023automated} have recently been proposed that eliminate the requirement of stress data in model training.
Instead, it leverages displacement and force data to learn the material's strain energy density function, enabling the formulation of hyperelastic constitutive laws directly from experimental observations. This method is especially beneficial in situations where stress measurements are unfeasible or susceptible to inaccuracies.
Of particular interest is the NN-Euclid \cite{thakolkaran2022nn} framework that utilizes the input constraint neural network (ICNN) to learn the constitutive relation from force-displacement data.
However, NN-Euclid \cite{thakolkaran2022nn} does not exploit known (but approximate) information about the material model, that often is a priori available. 
We hypothesize that exploiting such prior information can help the neural network to learn from sparse data and extrapolate better, and \textit{model-data fusion} is potentially the way forward in constitutive modelling.}

{In model-data fusion, the high-level idea is to integrate 
phenomenological and data-driven models \cite{fuhg2021model}. 
Along similar line, hybrid strategies already exist for combining data-driven models with physics-based models to solve partial differential equations (PDEs). For instance, Chakraborty \cite{chakraborty2021transfer} introduced a transfer learning-based multi-fidelity physics-informed deep learning framework, which demonstrated efficiency but lacked interpretability. Our recent works \cite{chakraborty2023dpa, chakraborty2023deep}, also proposed an alternative framework that augments known physics with deep learning models for solving PDEs and stochastic differential equations (SDEs). These hybrid frameworks are found to be accurate, more interpretable and generalize well to unseen environments.
However, contrary to these work, the goal in this paper is not to solve PDEs/SDEs using data-physics fusion; instead, the objective is to develop a hybrid framework for constitutive modelling utilising existing phenomenological models along with the deep learning frameworks.
Additionally, similar to \cite{thakolkaran2022nn}, we want to train the model using only the global reaction force and full-field displacement data.
This is crucial due to the difficulty in availing stress-strain data, the practical problem of data scarcity, the difficulty in formulating accurate phenomenological models, and the critical need for generalization when a material's behaviour changes under varying external conditions.
}

In this research work, we propose FuCe, a novel approach rooted in model-data augmentation for learning constitutive learning from global force displacement data. FuCe extends the NN-Euclid proposed in \cite{thakolkaran2022nn} by incorporating a known (but approximate) model into the overall framework. Therefore, all the advantages inherent in NN-Euclid also hold for FuCe while allowing learning from sparse data and better generalization beyond the training regime. 
Additionally, we also incorporate Monte Carlo-dropout (MC-dropout) into the architecture, making it uncertainty-aware. 
Overall, the salient features of the proposed FuCe approach are as follows:
\begin{itemize}
    \item \textbf{Model-data fusion:} The proposed framework utilizes classical phenomenological constitutive models alongside deep learning models, which augment and correct the phenomenological models to align with real material data.
    \item \textbf{Satisfies all necessary constitutive conditions:} Similar to NN-Euclid \cite{thakolkaran2022nn}, FuCe satisfies all thermodynamic and physical feasibility conditions.
    \item \textbf{Utilizes practically feasible force-displacement noisy data:} 
    As the proposed FuCe builds on top of NN-Euclid, 
    it allows learning from practically available full-field displacement and global reaction force data for training purposes, accounting for noise in the data.
    \item \textbf{Approximate Bayesian inference:} To quantify uncertainties in model predictions \cite{padmanabha2024improving}, a Bayesian approximation method called Monte Carlo Dropout \cite{gal2016dropoutbayesianapproximationrepresenting} is used during the inference stage. This makes the model uncertainty-aware and the uncertainty estimated can be used for decision-making, if needed.
\end{itemize}
Overall, the proposed framework is a novel hybrid model of its kind, that utilises deep learning models to learn the discrepancy between the known phenomenological model and the actual material behaviour utilizing the practically available full-field displacement and global reaction force data.

The remainder of the paper is organized as follows. The background of constitutive models specifically hyperelasticity is formally described
in Section \ref{sec:bck}. A detailed description of the problem statement is also presented at the end of  Section \ref{sec:bck}. The proposed approach along with the algorithm is described in Section \ref{sec:pp}. Results and discussion of the numerical experiments that are carried out are presented in Section \ref{sec:ns}. Finally, Section \ref{sec:concl} presents the
concluding remarks.

\section{Background and problem statement}\label{sec:bck}
\subsection{Stress-Deformation measures and Hyper-elasticity}\label{subsec:def}
Consider a solid body, having a reference configuration \(\mathbf{X} \subset \mathbb{R}^3\) at initial time  \(t_0 \in \mathbb{R}\). After deformation, at time \(t \in [t_0, \infty)\), the body attains its current configuration \(\mathbf{x}\subset \mathbb{R}^3\). This motion is described by a bijective mapping \(\boldsymbol{\zeta} : \mathbf{X} \times [t_0, \infty) \rightarrow \mathbf{x}\), during which a typical material particle in the reference configuration \(\mathbf{X}\) undergoes a displacement \(\mathbf{u}(\mathbf{X}, t)\) to arrive at their positions in the current configuration \(\mathbf{x}\), such that:
\begin{equation}
    \mathbf{x} = \boldsymbol{\zeta}(\mathbf{X}, t) = \mathbf{X} + \mathbf{u}(\mathbf{X}, t) \quad \text{with} \quad \mathbf{X} = \boldsymbol{\zeta}(\mathbf{X}, t_0)
\end{equation}
With regards to the aforementioned deformation, we define some deformation measures. At first, the gradient of the function $\boldsymbol{\zeta}(\mathbf{X}, t)$ is known as the deformation gradient $\bm{F} \in \mathcal{GL}^+(3)$ is defined as:
\begin{equation}
    \bm{F} = \nabla_{\mathbf{X}} \boldsymbol{\zeta} = \mathbb{I} + \nabla \mathbf{u}
\end{equation}
and its determinant as $ J = \frac{dV}{dV_0} = \det \bm{F} \in \mathbb{R}^+$, where \(\mathcal{GL}^+(3)\) denotes the set of all invertible second-order tensors with positive determinant. Another deformation measure, ensuring the absence of rigid body motions, is represented by the positive definite right Cauchy–Green deformation tensor:
\begin{equation}
    \mathbf{C} = \bm{F}^T  \bm{F}
\end{equation}
Lets now define some stress measures, starting with the true stress or Cauchy stress tensor $\bm{\sigma}$. It is a second-order symmetric tensor that describes the stress vector (a.k.a traction vector) $\mathbf{t}$ resulting from an infinitesimal force $d\mathbf{f}$ applied on an infinitesimal surface $d\mathbf{A}$ with normal $\mathbf{n}$ in the current configuration:
\begin{equation}\label{eq:stress}
\mathbf{t} = \lim_{d\mathbf{A} \to 0} \frac{d\mathbf{f}}{d\mathbf{A}} = \bm{\sigma} \mathbf{n}
\end{equation}
The infinitesimal force $d\mathbf{T}$ can be expressed either by taking area $d\mathbf{S}$ and its normal $\mathbf{n}$ considering the current configuration or by taking area $d\mathbf{S}_0$ and its normal $\mathbf{n}_0$ of the original (initial) configuration. This can be expressed as:
\begin{equation}
d\mathbf{f} = \bm{\sigma} \mathbf{n} \, d\mathbf{A} = \bm{P} \mathbf{n}_0 \, d\mathbf{A}_0
\end{equation}
where $\bm{P}$ is known as the (first) Piola-Kirchhoff stress tensor. The first Piola-Kirchhoff stress tensor $\mathbf{P}$ cam also be related to the Cauchy stress tensor $\bm{\sigma}$ through the deformation gradient $\bm{F}$ and the Jacobian $J$:
\begin{equation}
\mathbf{P} = J \sigma \bm{F}^{-T}
\end{equation}
Another measure of stress can be expressed by defining the undeformed force $d\mathbf{\tilde f} = \bm{F}^{-1} d\mathbf{f} $, which then can be defined as:
\begin{equation}
d\mathbf{\tilde f} = \bm{S} \mathbf{n_0} \, d\mathbf{A_0}
\end{equation}
where $\bm{S}$ denotes the second Piola-Kirchhoff stress tensor. It is a symmetric stress tensor that can be written as:
\begin{equation}
\mathbf{S} = J \bm{F}^{-1} \bm{\sigma} \bm{F}^{-T} = \bm{F}^{-1}\bm{P},
\end{equation}
with \(\rho_0\) and \(\rho\) representing the mass densities of the reference and current configuration, respectively, accounting for mass conservation, \(\rho_0 = J\rho\) yields the balance of linear momentum with regard to the reference configuration:

\begin{equation}
\nabla_{\mathbf{X}} \cdot \mathbf{P} + \rho_0 \mathbf{f} = 0.
\end{equation}
The preceding equation does not include inertia factors, and the mass-specific force density is indicated by \(\mathbf{f}\). Furthermore, the balance of angular momentum corresponds to:

\begin{equation}\label{eq:angmom}
  \mathbf{P} \cdot \mathbf{F}^T = \mathbf{F} \cdot \mathbf{P}^T,  
\end{equation}
when taking into account the balances of mass and linear momentum.

\subsection{General criteria for hyper-elasticity}\label{subsec:hyper}
In elasticity, the goal of constitutive modelling is to establish a relationship between the strain at a material location and the corresponding stress. However, in hyperelastic constitutive modelling, instead of a direct mapping, the stress-strain relationship is derived through a potential \(\psi \colon \mathcal{GL}^+(3) \to \mathbb{R}\), \(\bm{F} \to \psi(\bm{F})\), which corresponds to the strain energy density stored in the body. This relationship can be expressed as:

\begin{equation}\label{eq:Ppsi}
    \bm{P}(\bm{F}) = \frac{\partial \psi(\bm{F})}{\partial \bm{F}} \quad \text{and} \quad \mathbb{C}(\bm{F}) = \frac{\partial \bm{P}(\bm{F})}{\partial \bm{F}}, \quad \forall \,\bm{F} \in \mathcal{GL}^+(3)
\end{equation}
where \(\mathbb{C}(\bm{F})\) denotes the (incremental) tangent modulus. Thus, the goal of hyperelastic constitutive modelling is to learn this strain energy potential \(\psi(\bm{F})\). However, for any learned material model \(\psi(\bm{F})\) to be mathematically plausible, it must also meet the following thermodynamic and physical requirements:
\begin{itemize}
    \item \textbf{Thermodynamic consistency:} The stress tensor is thermodynamically consistent by construction since its definition as a gradient field in Eq. \eqref{eq:Ppsi} entails energy conservation and path independence \cite{kalina2023fe,linden2021thermodynamically, linka2023new}.

    \item \textbf{Stress tensor symmetry:} The construction of \(\psi(\bm{F})\) must guarantee consistency with the angular momentum balance. This requirement can be represented as follows by applying Eq. \eqref{eq:angmom} and \eqref{eq:Ppsi}:
\begin{equation}
   \frac{\partial \psi}{\partial \bm{F}} \cdot \bm{F}^T = \bm{F} \cdot \frac{\partial \psi}{\partial \bm{F}^T} 
\end{equation}
This leads to the necessity for the symmetry of the stress tensors \(\bm{\sigma}\) and \(\bm{S}\), using the stress transformations presented in Section \ref{subsec:def} for eg:
\begin{equation}
    \bm{S} = \bm{F}^{-1} \cdot \frac{\partial \psi}{\partial \bm{F}} \overset{!}{=} \frac{\partial \psi}{\partial \bm{F}^T} \cdot \bm{F}^{-T} = \bm{S}^T
\end{equation}

\item \textbf{Objectivity:} The constitutive model must be independent of the choice of material frame i.e:
\begin{equation}
  \psi(\bm{R} \bm{F}) = \psi(\bm{F}), \quad \forall \bm{F} \in \mathcal{GL}^+(3), \bm{R} \in \text{SO}(3),  
\end{equation}
where \(\text{SO}(3)\) denotes the 3D rotation group.

\item \textbf{Normalisation condition for stress and energy:}
In the absence of any deformation, the stress and the strain energy must be zero i.e.,
\begin{equation}
    \psi(\bm{F} = \bm{I}) = \bm{P}(\bm{F} = \bm{I}) = \mathbf{0}
\end{equation}
    
\item \textbf{Polyconvexity} 
The concept of polyconvexity in the calculus of variations is generalisation of the concept of quasicovexity. The material must be quasi-convex in order for it to be stable \cite{morrey1952quasi,schroder2010anisotropie}. Since, imposing quasi-convexity is not numerically tractable, this is relaxed by enforcing polyconvexity \cite{ball1976convexity,schroder2010anisotropie}. \(\psi(\bm{F})\) is considered polyconvex if there exists a convex function \(\mathcal{P}\) such that
\begin{equation}\label{eq:poly}
    \psi(\bm{F}) = \mathcal{P}(\bm{F}, \text{Cof} \bm{F}, \det \bm{F})
\end{equation}
\end{itemize}

\subsection{Classical Phenomenological models}\label{subsec:anamod}
In the study of hyperelastic materials, strain energy density functions are often defined using various deformation invariants such as the principal invariants of the right Cauchy-Green deformation tensor \( \bm{C} \): \( I_1 = \mathrm{tr}(\bm{C}) \), \( I_2 = \frac{1}{2} [ \mathrm{tr}(\bm{C})^2 - \mathrm{tr}(\bm{C}^2)] \), and \( I_3 = \det(\bm{C}) \). While most models were originally formulated with the assumption of incompressibility, it is crucial for constitutive models to account for compressibility, therefore we have assumed a deviatoric-volumetric split. The volumetric invariant \( J \) is given by the determinant of the deformation gradient \( \bm{F} \), specifically \( J = \det(\bm{F}) = I_1^{1/2} \). In isotropic materials, the deviatoric invariants are \( \tilde{I}_1 = J^{-2/3} I_1 \) and \( \tilde{I}_2 = J^{-4/3} I_2 \) and for anisotropic materials the deviatoric invariants are \( \tilde{I}_{\alpha_i} = J^{-2/3} (\mathbf{a}^{(i)} \cdot \bm{C} \mathbf{a}^{(i)}) \), where \( \mathbf{a}^{(i)} \) denotes the \( i^{th} \) fiber direction \cite{melly2021review, thakolkaran2022nn}. In case of unknown fiber orientations \( \{ \alpha_i : i = 1, 2, \ldots \} \), the fiber directions \( \mathbf{a}^{(i)} = (\cos \alpha_i, \sin \alpha_i, 0)^T \) are treated as trainable parameters. 

The invariants are further shifted by appropriate scalar values, and in some cases squared, to ensure that their values and the respective derivatives with respect to \( \mathbf{F} \) are zero in case of no deformation (\( \mathbf{F} = \mathbf{I} \)). In this work, we utilize specific forms of strain energy density functions from classical models for data generation and training purposes, as detailed below:

\begin{enumerate}
    \item \textbf{Neo-Hookean (NH) Model} \cite{rivlin1948large}:
    \begin{equation}\label{eq:neo}
        \psi(\mathbf{F}) = 0.5(\tilde{I}_1 - 3) + 1.5(J - 1)^2
    \end{equation}

    \item \textbf{Isihara (IH) Model} \cite{isihara1951theory} :
    \begin{equation}\label{eq:Is}
        \psi(\mathbf{F}) = 0.5(\tilde{I}_1 - 3) + (\tilde{I}_2 - 3) + (\tilde{I}_1 - 3)^2 + 1.5(J - 1)^2
    \end{equation}



    \item \textbf{Arruda–Boyce (AB) Model} \cite{arruda1993three}:
    \begin{equation}\label{eq:Ab}
         \psi(\mathbf{F}) = 2.5\sqrt{N_c}\left[\beta_c \lambda_c - \sqrt{N_c} \log\left(\frac{\sinh \beta_c}{\beta_c}\right)\right] - c_{AB} + 1.5(J - 1)^2
    \end{equation}
    
    where \(\lambda_c = \left(\frac{\tilde{I}_1}{3}\right)^{1/2}\), \(\beta = \mathcal{L}^{-1}\left(\frac{\lambda_c}{\sqrt{N_c}}\right)\), and \(\mathcal{L}^{-1}\) denotes the inverse Langevin function. Following \cite{thakolkaran2022nn}, the constants are set to \(N_c = 28\) (denoting the number of polymeric chain segments) and \(c_{AB} \approx 3.7910\), the latter used to offset the energy density to zero at \(\mathbf{F} = \mathbf{I}\) (since the Arruda–Boyce feature does not itself vanish for zero deformation).

    
   
    \item \textbf{Anisotropic Model with one fiber family at \(\alpha_1 = 45^\circ\) orientation (AI45)}:\label{eq:ani45}
    \begin{equation}
          \psi(\mathbf{F}) = 0.5(\tilde{I}_1 - 3) + 0.75(J - 1)^2 + 0.5(\tilde{I}_{\alpha_1} - 1)^2
    \end{equation}


   
\end{enumerate}
These models struggle to capture the complex behaviours of real materials due to their fixed form and restrictive functional dependence on invariants. The difficulty further escalates when trying to meet all the physical constraints outlined in Section \ref{subsec:hyper}. Additionally, each model has specific limitations under varying deformation levels and loading conditions. For example, the Neo-Hookean model is advantageous due to its simplicity and computational efficiency; however, it requires stringent conditions to achieve accurate results. The Mooney-Rivlin model performs well for small to medium strains, while the Gent model excels in predicting large strain loadings (up to 300\%). The Isihara model is accurate for moderate stretch values but poorly estimates behaviour under biaxial loading. The Ogden model is highly suitable for large deformation behaviour; however, its drawback is the need for different sets of material parameters for each deformation mode.

Addressing these limitations can be achieved through the use of machine learning models. Traditional models often require extensive stress-strain data, which is not always practical. It is crucial to train the model using realistically measurable full-field displacement and global reaction force data. End-to-end training with force-displacement data is challenging and necessitates a large dataset. Furthermore, these models struggle to generalize beyond the training regime. A logical step forward is to integrate existing phenomenological models with the data driven models. This hybrid model can train effectively with less data and generalize well for large strains, overcoming the limitations of the traditional methods.

\subsection{Problem statement}\label{sec:ps}
{Material characterisation involves understanding the constitutive behaviour of the material under various loading conditions. In practice, reaction forces and displacement measurements are often used as primary data, however, these measurements are typically sparse and are susceptible to inaccuracies. This poses challenges in accurately discovering material behaviour. On the other hand, phenomenological models are often approximate and fail to capture the material's true response accurately.
We hypothesize that these challenges can be addressed by augmenting known phenomenological models with machine learning-based models. 
}

To formally define the problem, consider a material domain \(\Omega\) undergoing quasi-static loading in displacement-controlled testing. The configuration incorporates Dirichlet and Neumann boundary conditions, with the applied forces corresponding to reaction forces at constrained boundaries. The data comprises limited displacement measurements, and net reaction forces collected over load steps on specific boundary segments, emulating load cell measurements. With this setup, the objective is to discover the underlying constitutive law of the unknown material. Note that \cite{thakolkaran2022nn} have already tackled this problem setup; however, the key difference here resides in the fact that we have prior (approximate) knowledge that the material exhibits hyperelastic behaviour, and a potential constitutive model is also available. Since the constitutive model is approximate, the obtained stress field will also be inaccurate, leading to an erroneous force field solution. With this setup, the objective is to learn a hybrid model that fuses the known phenomenological model with the data-driven neural network.

\section{Proposed approach}\label{sec:pp}
In this section, we will present the detailed architecture of the proposed framework. This includes the integration of model and data, the design of the physics-based neural network architecture, and the implementation of correction terms to ensure compliance with the physical constraints outlined in Section. \ref{subsec:hyper}. Additionally, we will describe the algorithm, provide a flow diagram, discuss the loss function, elaborate on the ensembling and Bayesian inference methods, and detail the training process.

\subsection{Architecture design}
In a realistic setting, while developing a constitutive model for hyperelastic materials, we generally have access to a low-fidelity phenomenological model and sparse data. In this work, we propose a framework that integrates both sources of information. The fundamental idea is to augment the partially correct phenomenological model (base phenomenological model) with a data-driven deep learning architecture, which can be expressed as:

\begin{equation}\label{eq:aug_eqn_p}
     \psi(\bm{F}) = \psi_{\text{known}}(\bm{F}) + \psi_{NN}(\bm{F};\bm{\theta})
\end{equation}
Here, \(\psi_{NN}(\bm F;\theta)\) represents the neural network component, which is responsible for correcting \(\psi_{\text{known}}(\bm F)\) (the known phenomenological model). The hypothesis is that the neural network will learn the $\psi_{\text{missing}}(F)$, thereby improving the accuracy of the model. We will exploit the augmented strain energy density function \(\psi(\bm{F})\) to derive the stress measures, specifically the first Piola-Kirchhoff stress \(\bm{P}(\bm{F})\) and the incremental tangent modulus \(\mathbb{C}(\bm{F})\), as described by Eq. \eqref{eq:Ppsi}. Therefore, the primary objective is to accurately determine the function \(\psi(\bm{F})\). However, this augmented model \(\psi(\bm{F})\) must also satisfy all the physical and thermodynamic constraints outlined in Section.\ref{subsec:hyper}.

\subsubsection{Fundamental conditions}
The thermodynamic consistency is automatically ensured by constructing the stress measures as the gradient of a potential (strain energy density), as defined in Eq. \eqref{eq:Ppsi}. Additionally, if the strain energy density potential is formulated in terms of the invariants of the right Cauchy-Green deformation tensor \(\bm{C} = \bm{F}^T \bm{F}\), or equivalently, the Green-Lagrange strain tensor \(\bm{E} = (\bm{F}^T \bm{F} - \bm{I}) / 2\), denoted as \(\psi(\bm{E}(\bm{F}))\), the resulting Cauchy stress tensor is symmetric and inherently satisfies the condition of objectivity. As shown in Section. \ref{subsec:anamod}, the known phenomenological model is formulated in terms of invariants. Therefore, we also design the network with these same invariants as input to the network. With this setup, the model automatically satisfies the first three fundamental conditions: thermodynamic consistency, objectivity, and the symmetry of the stress tensor.

\subsubsection{Polyconvexity}\label{subsubsec:poly}
In order to ensure the condition of polyconvexity of the strain energy density potential, we have adopted Input Convex Neural Network (ICNN) \cite{amos2017input} as the neural network architecture. In these Neural Networks, the output is convex with respect to its input arguments. This is ensured by using convex and non-decreasing activation functions and non-negative weights. The Non-negative weights result in convex linear transformations in the hidden layers, as non-negative weighted sums of convex functions remain convex. Additionally, non-decreasing activation functions preserve convexity by ensuring that applying a non-decreasing convex function to a convex input results in a convex output, thereby ensuring the ICNN output is convex in the strain invariants. Further details on the ICNN employed in this work are provided in Section. \ref{subsec:icnn}.

In addition to the specialized architecture, ensuring polyconvexity also requires that the considered invariants themselves be polyconvex. However, in \(\psi(\bm E(\bm F))\), \(\bm E\) is not always convex in \(\bm F\), making it challenging to ensure polyconvexity in \(\bm F\) using Eq. \eqref{eq:poly}. Although, it is observed that the convexity of \(\psi\) in \(\bm E\) implies local convexity in \(\bm F\), which ensures local material stability \cite{yang2017convexity, as2022mechanics}. Note that local convexity is preferable over global convexity because it ensures stability without requiring the strain energy density to remain smooth and bounded as \(\det(F) \to 0^+\). This flexibility allows the strain energy density to approach infinity, implicitly fulfilling the growth condition necessary for physical realism under extreme compressions. In contrast, global convexity would prevent this by mandating a bounded energy response even in these extreme cases and would require explicitly satisfying the growth condition by adding correction terms \cite{thakolkaran2022nn}. Therefore, we also limit the architecture to local convexity only.

\subsubsection{Normalisation conditions}
In the undeformed configuration (\(\mathbf{F} = \mathbf{I}\)), it is essential that both the energy and stress normalization conditions are satisfied. This is achieved by incorporating physics-based terms to ensure the energy density vanishes at zero deformation. To achieve energy normalization, we add \(\psi_{energy}\) such that
\begin{equation}
   \psi(\bm{F})\big|_{\bm{F}=\bm{I}} = \left\{\psi_{\text{known}}(\bm{F}) + \psi_{NN}(\bm{F}; \bm{\theta})\right\}_{\big|_{\bm{F}=\bm{I}}} = 0
\end{equation}
The known phenomenological models are constructed to automatically satisfy the normalization condition, i.e., \(\psi_{\text{known}}(\mathbf{I}) = 0\). Consequently, the energy term is defined as:
\begin{equation}\label{eq:energycor}
    \psi_{energy} = -\psi_{NN}(\bm{I}; \bm{\theta})
\end{equation}

Furthermore, the stress correction term $\psi_{\text{stress}}$ is included in the potential to ensure stress vanishes under zero deformation. This term is defined as:
\begin{equation}\label{eq:stresscor}
    \psi_{\text{stress}} = \bm{H} : \bm{E}
\end{equation}
Utilizing Eq. \eqref{eq:Ppsi}, we have:
\begin{equation}
    \bm{P}(\bm{I}) = 0 = \left\{\frac{\partial \psi_{NN}(\bm{F}; \bm{\theta})}{\partial \bm{F}} + \bm{F} \bm{H}\right\}_{\big|_{\bm{F}=\bm{I}}} \implies \bm{H} = \frac{\partial \psi_{NN}(\bm{F}; \bm{\theta})}{\partial \bm{F}}\bigg|_{\bm{F}=\bm{I}}
\end{equation}
where $\bm{H}$ is a constant symmetric matrix. Here, we exploit the fact that  $\psi_{\text{known}}(\bm{F}) = 0$ for $\bm{F} = \bm{I}$ implies its derivative with respect to $\bm{F}$ at $\bm{F} = \bm{I}$ is also zero. Consequently, we obtain the following expression for the strain energy density:
\begin{equation}\label{eq:aug_eqn}
     \psi(\bm{F} ;\bm{\theta)} = \psi_{\text{known}}(\bm{F}) + \psi_{NN}(\bm{F};\bm{\theta}) + \psi_{\text{energy}} +  \psi_{\text{stress}}
\end{equation}
This is referred to as `augmented strain energy' and is used in this work.

\subsection{ICNN Architecture}\label{subsec:icnn}
Having discussed the fundamental constraints and a systematic procedure for satisfying the same, we proceed to discussing the architecture of the Input Convex Neural Network (ICNN), which is a key component of the proposed approach. Similar to previous work \cite{thakolkaran2022nn}, the inputs to the ICNN are modified versions of various invariants of \(\bm{C}\), as detailed in Section.\ref{subsec:anamod}. The architecture consists of repeated convex layers, each with 64 neurons in the hidden layer. It incorporates skip connections from the input to the hidden layer after each convex layer. This sequence is repeated three times. Convexity is ensured by maintaining positive weights, achieved by applying the softplus function to the weight matrix prior to its use, as shown below:
\begin{equation}
    \bm{z}^{l} = \sigma_1 \left(
    \underbrace{\sigma_2 \left( \bm{W}^{l} \right) \bm{z}^{l-1}}_{\text{convex-linear}} + 
    \underbrace{\bm{W}_s^{l} \bm{z}^{0}}_{\text{skip-connect}} + 
    \underbrace{\bm{b}^{l}}_{\text{bias}}
    \right)
\end{equation}
where $\bm{W}^{l} \in \mathbb{R}^{d^l \times d^{l-1}}$, $\bm{W}_s^{l} \in \mathbb{R}^{d^l \times |\bm{z}^0|}$ and $\bm{b}^{l} \in \mathbb{R}^{d^l}$ are trainable parameters with $d^l$ being the dimension of $l^{th}$ hidden layer, and
\begin{equation}
    \sigma_1 (z) = \underbrace{\mathcal{\gamma}_1 \left( \log(1 + e^z) \right)^2}_{\text{squared-softplus}}, \quad \text{and} \quad \sigma_2(z) = \underbrace{\mathcal{\gamma}_2 \left( \log(1 + e^z) \right)}_{\text{softplus}} 
\end{equation}  
respectively, for all \(z \in \mathbb{R}\). Here, \(\mathcal{\gamma}_1 > 0\) and \(\mathcal{\gamma}_2 > 0\) are hyperparameters.
{The softplus function ensures \(C^{\infty}\) continuity of the strain energy density and its derivatives, providing smooth and continuous stress and tangent operators. We consider the function \(\sigma_1\) as the squared softplus to handle the vanishing second derivative problem. This squaring not only helps to compute the tangent of the strain energy density but also stops disappearing gradients during backpropagation in neural network training (see \cite{goodfellow2016deep}).}
Additionally, skip connections are incorporated to mitigate the issues of vanishing gradients, overfitting, and accuracy degradation (or saturation) as the neural network architecture deepens, see reference \cite{he2016deep}. At the output of the neural network we get the missing component of the strain energy density.
\subsection{Training}


Consider we have nodal displacement data represented as \( u^t(r,j) \), where \( t = \{1, 2, \ldots, n_t\} \) denotes the load step, \( r = \{1, 2, \ldots, n_d\} \) indexes the nodal points, and \( j = \{1, 2\} \) indicates the direction of displacement. We define the degree of freedom at the \( r^{\text{th}} \) node in the \( j^{\text{th}} \) direction as \( \mathcal{D}_f = (r,j) \). This encompasses both \( \mathcal{D}_f^{\text{free}} \): degrees of freedom within the boundary where displacement is not fixed, and \( \mathcal{D}_f^{\text{fixed}} \): degrees of freedom at the boundaries where displacement is constrained. Additionally, we have reaction force data \( R_{\beta} \) available at selected boundary nodes \( \beta = 1, 2, \ldots, n_{\beta} \).

Degrees of freedom corresponding to these $n_{\beta}$ boundary nodal points constitute $\mathcal{D}_f^{fixed}$. 
\begin{figure}[ht!]
    \centering
\includegraphics[width=0.4\textwidth]{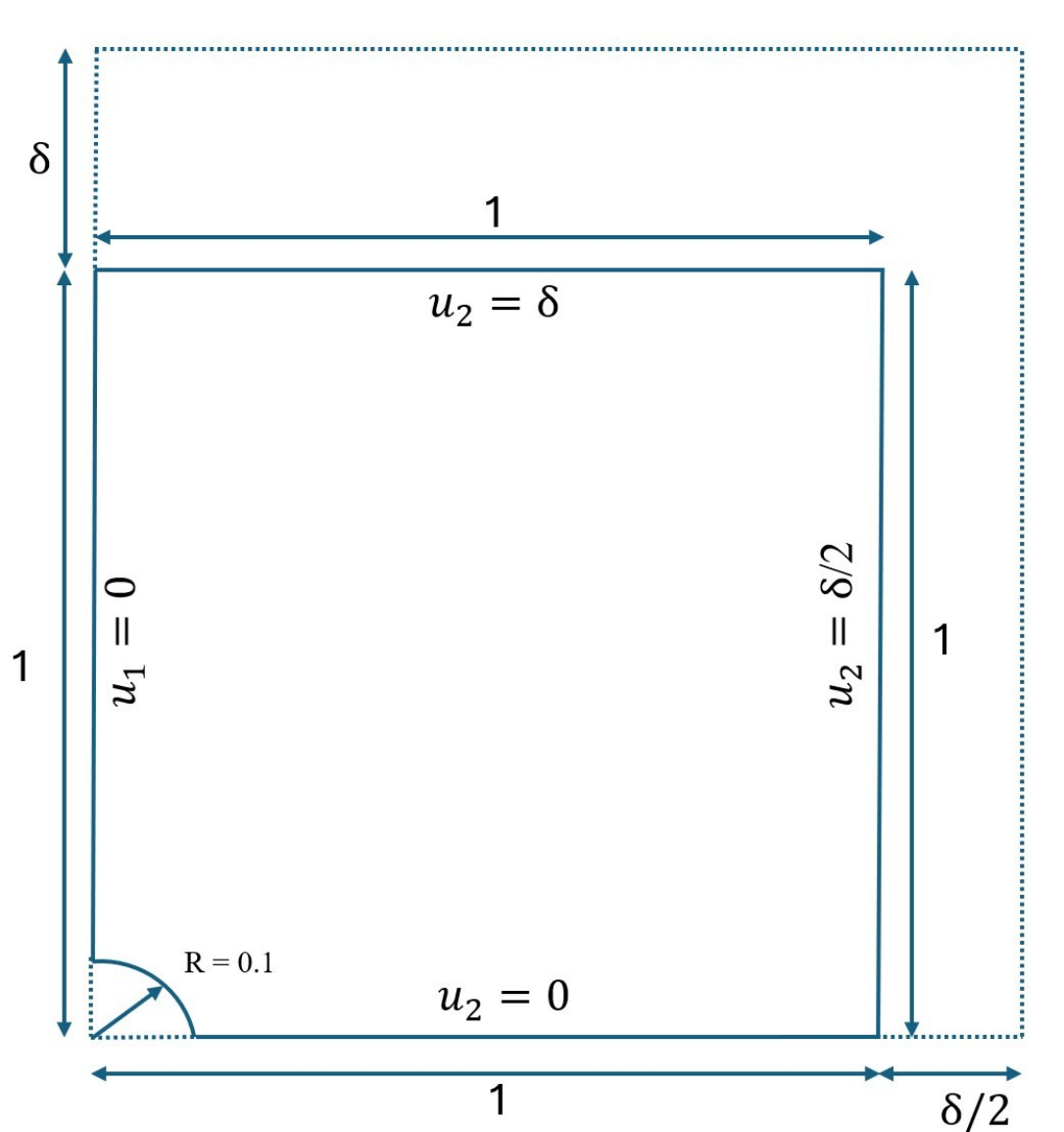}
    \caption{Training Setup: A square plate with a hole in the bottom-left corner is subjected to asymmetric biaxial tension under displacement control. The noisy data collected from full-field displacements and reaction forces is used to train the proposed constitutive model. This setup is the same as \cite{thakolkaran2022nn} so as to illustrate the advantage of model-physics fusion.}
    \label{fig:train_geom}
\end{figure}
Next, we divide the geometry into finite elements and write the displacement as a continuous function of spatial coordinates using shape functions \(\{N_r(X) : r = 1, \ldots, n_d\}\),
\begin{equation}\label{eq:ushp}
   \bm{u}(\bm{X}) = \sum_{r=1}^{n_d} N_r(\bm{X}) \bm{u}^t_{r}
\end{equation}
and \(N_r : \Omega \rightarrow \mathbb{R}\) is  associated with the node at reference coordinates \(X = \{X_r \in \Omega : r = 1, \ldots, n_d\}\). The deformation gradient is then given by:
\begin{equation}\label{eq:defg}
    \bm{F}(\bm{X}) = \bm{I} + \sum_{r=1}^{n_d} \bm{u}_r \otimes \nabla N_r(\bm{X})
\end{equation}
where \(\nabla\) denotes the gradient operator with respect to the reference coordinates. 
Similarly, let's introduce the test function $\bm{v}$ and describe it in terms of shape functions as:
\begin{equation}\label{eq:vshp}
    \bm{v}(\bm{X}) = \sum_{r=1}^{n_d} N_r(\bm{X}) v^t_{r,j} \quad \text{with} \quad v^t_{r,j} = 0 \quad \forall (r, j) \in \bigcup_{\beta=1}^{n_{\beta}} \mathcal{D}_f^{fixed}
\end{equation}
From the displacement field $\bm{u}(\bm{X})$, we obtain the deformation gradient following Eq.~(\ref{eq:defg}), modified invariants of which are input into the ICNN to yield the strain energy density potential. Using Eq.~(\ref{eq:Ppsi}), we derive the stress field. In the absence of stress field data, we use physics-informed loss in the form of the linear momentum conservation equation to inform and guide the training process. Note that although we use the finite element method in this work, other techniques including finite volume and finite difference can also be used.

In the reference domain $\Omega$, the weak form of the linear momentum balance under quasi-static loading (negligible inertia and body forces) is given by:
\begin{equation}\label{eq:linmom}
    \int_{\Omega} \mathbf{P} : \nabla \mathbf{v} \, dV - \int_{\partial \Omega_t} \hat{\mathbf{t}} \cdot \mathbf{v} \, dS = 0 \quad \forall \text{ admissible } \mathbf{v},
\end{equation}
where $\mathbf{v}$ is an admissible test function, sufficiently regular and vanishing on the Dirichlet boundary $\partial \Omega_u$, and $\hat{\mathbf{t}}$ represents the surface traction on $\partial \Omega_t$. In displacement-controlled testing, $\hat{\mathbf{t}} = 0$. The weak form is preferred over the strong form due to its reduced sensitivity to noise in the data. 
Utilizing Eqs.~(\ref{eq:vshp}) and (\ref{eq:linmom}), we obtain:
\begin{equation}\label{eq:intmom}
    \sum_{r=1}^{n_d} v_{r,j} \mathcal{F}_{r,j} = 0, \quad \text{where} \quad \mathcal{F}_{r,j} = \underbrace{\int_{\Omega} P_{jk} \nabla_k N_r \, dV}_{\text{internal force}} - \underbrace{\int_{\partial \Omega_t} \hat{t}_j N_r \, dS}_{\text{external force}}
\end{equation}
Here, $\mathcal{F}_{r,j}$ is defined as the difference between the external force due to applied tractions on Neumann boundaries and the interior force from the constitutive model. Numerical quadrature over the mesh is used to compute the integrals in Eq.~(\ref{eq:intmom}).
By simplifying Eq.~(\ref{eq:intmom}) separately for fixed and free degrees of freedom, we obtain:
\begin{equation}\label{eq:preloss}
    \mathcal{F}_{r,j} = 0 \quad \forall (r, j) \in \mathcal{D}_f^{\text{free}} \quad \text{and} \quad \sum_{(r,j) \in \mathcal{D}_f^{\text{fixed}}} \mathcal{F}_{r,j} = R_{\beta} \quad \forall \, \beta = 1, \ldots, n_{\beta}
\end{equation}
Although we have used finite element implemented using a differentiable finite element simulator, the overall approach can be seamlessly integrated with other numerical techniques such as finite volume method and finite difference method. {For more detials about the Eq.~(\ref{eq:preloss}) and Eq.~(\ref{eq:intmom}) interested readers can refer to \cite{thakolkaran2022nn}.}

To optimize the learnable parameters, we {employ a loss function similar to the one used in \cite{thakolkaran2022nn}, which can be expressed as:}
\begin{equation}\label{eq:loss}
    \mathcal{L} = \sum_{t=1}^{n_l} \left[\sum_{(r,j) \in \mathcal{D}_f^{\text{free}}} (\mathcal{F}^t_{r,j})^2 + \sum_{(r,j) \in \mathcal{D}_f^{\text{fixed}}} (\mathcal{F}^t_{r,j} - R_{\beta})^2\right]
\end{equation}
Note that unlike \cite{thakolkaran2022nn}, our overall model also includes the known phenomenological model. Note that although 
{the network (corresponding to the missing physics) exhibits convexity with respect to its input but is non-convex in relation to the trainable parameters. Moreover, due to the absence of direct stress or strain energy labels, we utilize global reaction force data for training.}
These factors lead to an optimization problem with multiple solutions (multiple local minima), making the solution highly sensitive to initialization. To address this challenge, we trained multiple networks independently with different random initialisations. 
{The model with the lowest loss value among all the trained candidates is selected and subjected to Monte Carlo dropout during prediction to account for uncertainty. The details of the training process are described in Algorithm \ref{alg:training}.}

\begin{algorithm}[!]
\caption{Training Algorithm for FuCe}\label{alg:training}
\SetAlgoLined

\KwIn{Displacement data \(U = \{\bm{u}^t_{r,j}\},\) where \(r = 1, \ldots, n_d,\) \(j = 1, 2,\) \(t = 1, \ldots, n_l\); and global reaction forces \(R^{\beta} = \{R^t_{r,j}, \beta = 1, \ldots, n_i, t = 1, \ldots, n_t\}\)}

\KwOut{Trained hybrid model \(\psi(\bm{F}; \bm{\theta^*})\) with optimised parameters $\bm{\theta^*}$}

\textbf{Initialize} ICNN parameters \(\bm\theta\) and Adam optimizer\;

\For{$e = 1, \ldots, epochs$}{
  \textbf{Set} $\mathcal{L} \gets 0$ \tcp*{Initialize epoch loss}
  
  \For{$t = 1, \ldots, n_l$}{
    \For{each element in mesh}{
      Compute energy and stress correction terms \(\psi_{energy}\), \(\psi_{\text{stress}}\) \tcp*{see \ref{eq:energycor}, \ref{eq:stresscor}}
      Update \(\psi(\bm{F}; \bm{\theta})\) and \(\bm{P}(\bm{F}; \bm{\theta})\) \tcp*{see \ref{eq:aug_eqn}}
    }

    Compute forces \(\mathcal{F}^t_{r,j}\) for all $r, j$ \tcp*{see \ref{eq:intmom}}
    Update loss \(\mathcal{L} \) for free degrees of freedom \((r, j) \in \mathcal{D}_f^{free}\) \tcp*{see \ref{eq:preloss}}
  }

  \For{$\beta = 1, \ldots, n_{\beta}$}{
    Compute reaction forces \(r^{\beta, t}\) for fixed degrees of freedom \((r, j) \in \mathcal{D}_f^{fixed}\)\;
    Update \(\mathcal{L} \) using \((R^{\beta} - r^{\beta, t})^2\)\;
    \(\mathcal{L} \gets \mathcal{L} + \left(R^{\beta} - r^{\beta, t}\right)^2\)\tcp*{see \ref{eq:preloss} }}
    }

  \textbf{Update} \(\theta\) using Adam optimizer and gradients \(\partial \mathcal{L} / \partial \theta\)\;
\end{algorithm}

After selecting the best-converged network, it is still important to quantify the uncertainty in the predictions. To achieve this, we used approximate Bayesian inference during the prediction stage by employing the Monte Carlo dropout method. Details on the same is provided in the next section.
\subsection{Approximate Bayesian inference using Monte Carlo dropout}
Bayesian approaches offer a powerful framework by providing both predictions and their associated uncertainty. Rather than employing complex Bayesian methods, we utilize Monte Carlo dropout, which is a simpler yet effective approximate Bayesian technique for capturing predictive uncertainty. This approach uses dropout during predictions and runs several forward passes with different dropout masks, which results in producing a distribution of predictions.

Consequently, $N_m$ dropout masks are applied to the best-trained model, generating $N_m$ (taken as $100$) number of outputs for strain energy density $\hat{\psi}(\bm{F};\bm{\theta})$ and derived stress measure Piola-Kirchhoff stress $\bm{\hat{P}}(\bm{F};\bm{\theta})$ , denoted as $\{\hat{\psi}_1, \hat{\psi}_2, \dots, \hat{\psi}_{N_m}\}$ and $\{\hat{\bm{P}}_1, \hat{\bm{P}}_2, \dots, \hat{\bm{P}}_{N_m}\}$. The predicted outputs for a test input $\bm{F}$ during the $n^{\text{th}}$ forward pass of dropout can be expressed as:
\begin{equation}
\hat{\psi_n} = \psi(\bm{F}; \bm{\theta}_n^*,D_n) \quad \text{and} \quad \hat{\bm P_n} = \frac{\partial \psi(\bm{F}; \bm{\theta}_n^*,D_n)}{\partial \bm{F}}
\end{equation}
where $n = \{1, 2, \dots, N_m\}$ and $\bm \theta_n^*$ represents the parameters of the trained model after applying $n^{\text{th}}$ dropout mask, represented by $D_n$, each element of which is a Bernoulli random variable. After obtaining the $N_m$ predicted outputs, the expectation of the predicted outputs is estimated as:
\begin{equation}
\mathbb{E}[\hat{\psi}] = \frac{1}{N_m} \sum_{n=1}^{N_m} \hat{\psi_n} \quad \text{and} \quad \mathbb{E}[\hat{\bm P}] = \frac{1}{N_m} \sum_{n=1}^{N_m} \hat{\bm P_n}
\end{equation}

\begin{figure}[ht!]
    \centering
\includegraphics[width=1\textwidth]{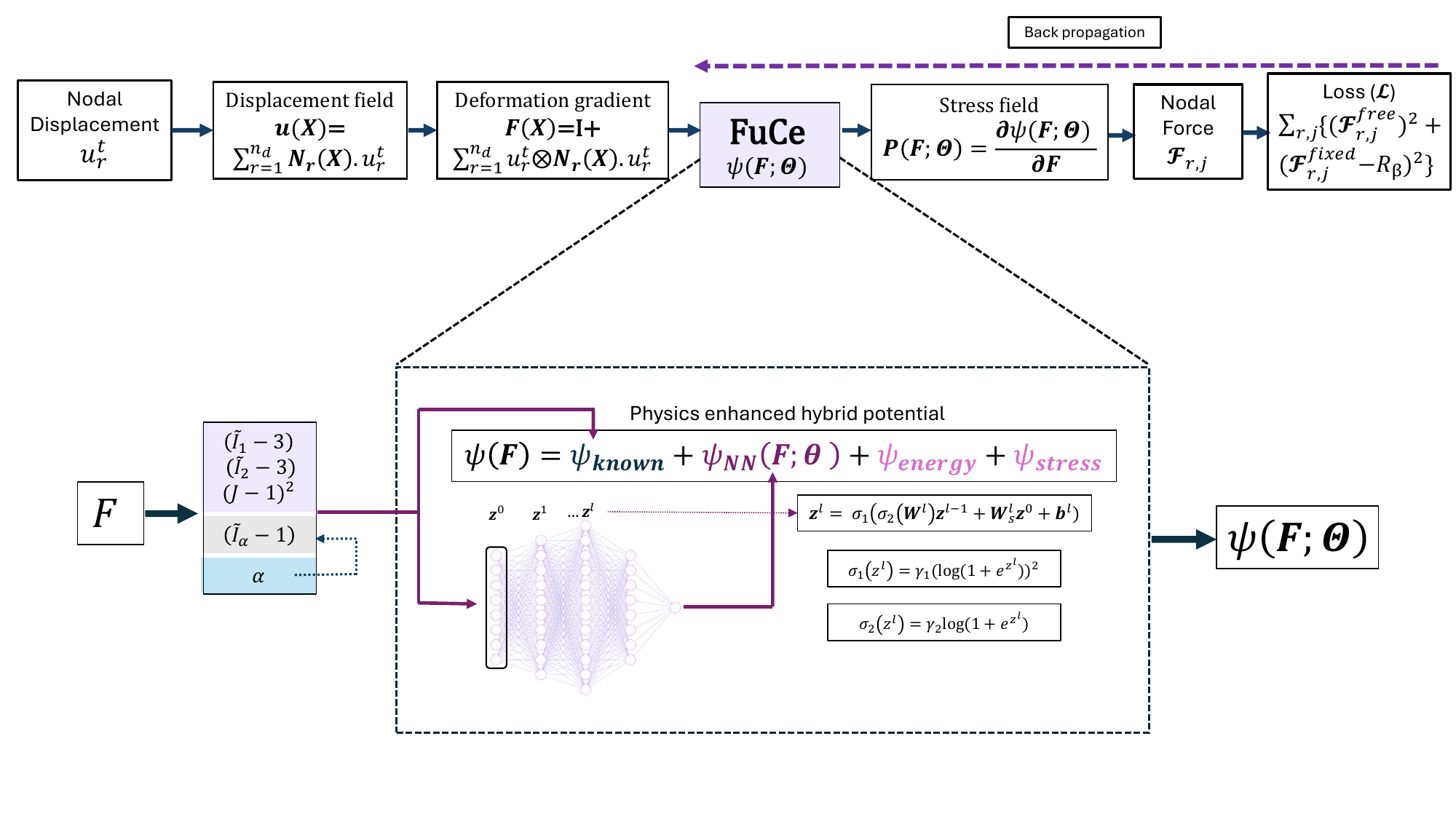}
    \caption{The overall framework of the FuCe. Note that for training the proposed approach, differentiable physics simulator is needed that allows backpropagation through the numerical solver.}
    \label{fig:overall}
\end{figure}

We then calculate the empirical variance of these outputs to estimate the predictive uncertainty of the model as follows:
\begin{equation}
\operatorname{Var}(\hat{\psi_n}) = \frac{1}{N_m-1} \sum_{n=1}^{N_m} (\hat{\psi_n} - \mathbb{E}[\hat{\psi_n}])^2 \quad \text{and} \quad \operatorname{Var}(\hat{\bm P}) = \frac{1}{N_m-1} \sum_{n=1}^{N_m} (\hat{\bm P_n} - \mathbb{E}[\hat{\bm P_n}])^2
\end{equation}
We quantify the model's epistemic uncertainty by calculating the empirical variance of these outputs, enhancing the trustworthiness of the model's predictions, particularly in scenarios with sparse data. This method strengthens the predictive capability of our framework and aligns with the increasing demand for models that offer predictive uncertainty alongside predictions, thereby supporting more informed decision-making.
The overall framework proposed in this work is shown in Fig. \ref{fig:overall}.

\section{Results and Discussion}\label{sec:ns}
In this section, we demonstrate the effectiveness of our approach through various numerical examples. We investigate the effectiveness of the proposed approach in learning the constitutive relation from sparse and noisy data. Additionally, we also showcase its practical utility by fusing the learning hybrid model with finite element solver \texttt{FeNICS} \cite{AlnaesEtal2015} 

\subsection{Simple Stress States}
In this section, we present a comparative evaluation of the proposed data-model fusion framework, the known phenomenological model, and a purely data-driven model without fusion (NN-Euclid), all benchmarked against the ground truth. The performance of these models is assessed across six stress states: simple shear (SS), pure shear (PS), uniaxial tension (UT), uniaxial compression (UC), biaxial tension (BT), and biaxial compression (BC), with the corresponding deformation gradients provided below:
\begin{equation}\label{eq:defgrad}
\begin{aligned}
\mathbf{F}^{\text{UT}}(\gamma) &= \begin{bmatrix}
1 + \gamma & 0 \\
0 & 1
\end{bmatrix}; \quad
\mathbf{F}^{\text{UC}}(\gamma) = \begin{bmatrix}
\frac{1}{1+\gamma} & 0 \\
0 & 1
\end{bmatrix}, \\
\mathbf{F}^{\text{BT}}(\gamma) &= \begin{bmatrix}
1 + \gamma & 0 \\
0 & 1 + \gamma
\end{bmatrix}, \quad
\mathbf{F}^{\text{BC}}(\gamma) = \begin{bmatrix}
\frac{1}{1+\gamma} & 0 \\
0 & \frac{1}{1+\gamma}
\end{bmatrix}, \\
\mathbf{F}^{\text{SS}}(\gamma) &= \begin{bmatrix}
1 & \gamma \\
0 & 1
\end{bmatrix}, \quad
\mathbf{F}^{\text{PS}}(\gamma) = \begin{bmatrix}
1 + \gamma & 0 \\
0 & \frac{1}{1+\gamma}
\end{bmatrix}.
\end{aligned}
\end{equation}
where $\gamma$ represents the loading parameter. Note that these deformation paths are used only during prediction for evaluating the model. During training, a single model is trained based on complex generalised stress states, rather than on individual deformation paths. We have utilized FEM-emulated digital image correlation data from 
\cite{thakolkaran2022nn}. To emulate a realistic scenario, the dataset is corrupted with Gaussian white noise. {Further, to illustrate the capability of the proposed approach in handling small and sparse dataset, we only utilized 50\% of the load steps originally used in \cite{thakolkaran2022nn}.} 
We evaluate the performance of the proposed constitutive framework using three different elastic solids, as detailed in Section \ref{sec:Is}, Section \ref{sec:Ab}, and Section \ref{sec:An45}. To assess the model's extrapolative capabilities, $\gamma$ is extended beyond its training range of $\gamma \in [0,1]$. Furthermore, when employing the approximate Bayesian method to predict the mean output and confidence interval, a dropout rate of $0.5$ is used during Monte Carlo dropout. 

\subsubsection{Isihara}\label{sec:Is}
%
In this case, we have chosen the Isihara hyperelastic model as our ground truth. Originally formulated by Isihara et al. \cite{isihara1951theory}, this model is renowned for its high accuracy in predicting the response of hyperelastic materials. The enhanced accuracy is attributed to the incorporation of the second invariant in the formulation of the strain energy function, as presented in Eq. \eqref{eq:Is}. We considered the Neo-Hookean model (Eq.~\eqref{eq:neo}) as the known phenomenological model $\psi_{\text{known}}$ and trained a single deep learning model $\psi_{NN}$ on the generalised state of stress (with only four load steps) to learn the discrepancies. 
To enable training the model directly from displacement measurements, we developed a differentiable physics solver that allows backpropagation through the solver. This allows us to train the model directly from the displacement data. The model is trained based for \(\gamma \in [0, 1]\); however, during evaluation, we considered \(\gamma \in [0, 3]\) to investigate the extrapolation performance of the model. Additionally, the trained model is evaluated in six different states of stress mentioned earlier. 

\begin{figure}[ht!]
    \centering
    \begin{subfigure}{0.5\textwidth}
        \centering
        \includegraphics[width=\textwidth]{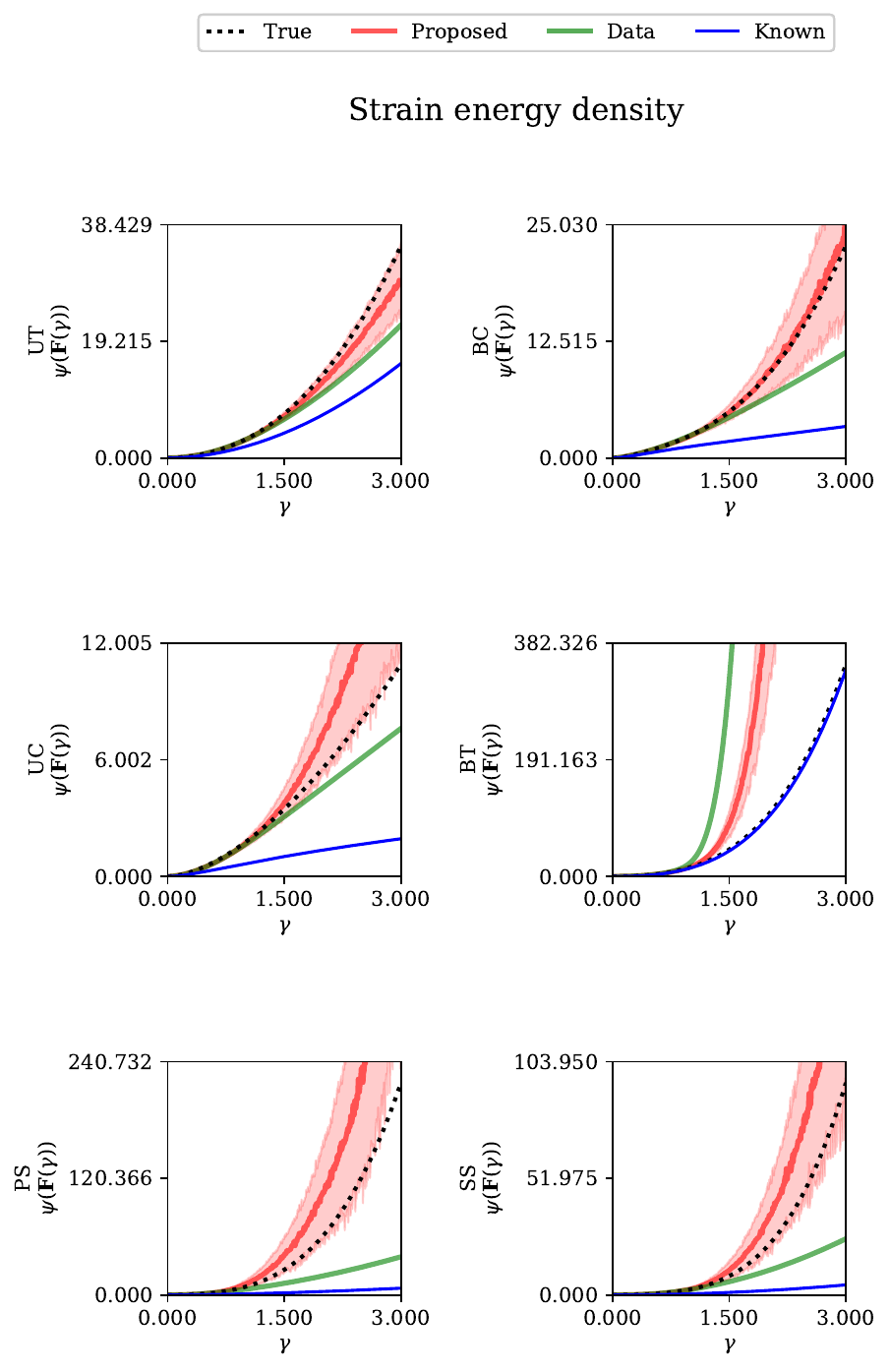}
        \caption{}
        \label{fig:se_n_i}
    \end{subfigure}%
    \begin{subfigure}{0.5\textwidth}
        \centering
        \includegraphics[width=\textwidth]{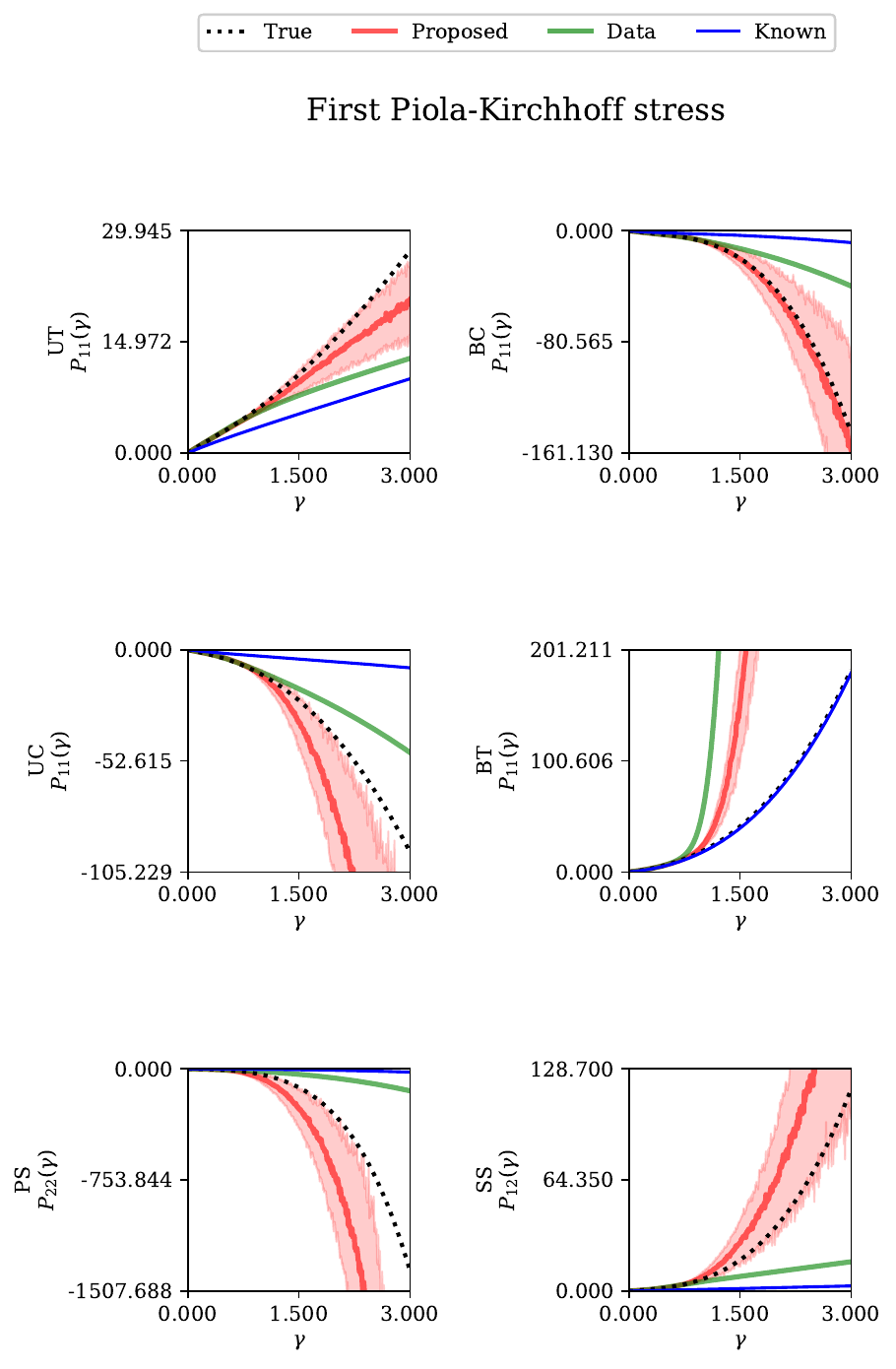}
        \caption{}
        \label{fig:pks_n_i}
    \end{subfigure}
    \caption {Comparison of predicted outputs: (a) Strain energy density \(\psi(F(\gamma))\) and (b) First Piola–Kirchhoff stress \(\bm{P}(\bm{F}(\gamma))\), produced by the proposed constitutive model, against a data-driven model without fusion (NN-Euclid) and a known phenomenological model, relative to the ground truth across six distinct stress states. The Neo-Hookean model is utilized as the known phenomenological model, along with the known fibre direction \(\alpha\), while the Isihara model serves as the ground truth.}
    \label{fig:n_i}
\end{figure}

Figures \ref{fig:se_n_i} and \ref{fig:pks_n_i} illustrate the variation in strain energy density and the Piola-Kirchhoff stress tensor, respectively, as functions of the loading parameter $\gamma$ for six deformation paths mentioned above. The proposed framework consistently outperforms the other models across all deformation paths, except for BT.
One potential solution towards improving the prediction for BT is to fine-tune the model for BT data; however, the same is not explored in the work. 
We note that purely data-driven models also perform well for $\gamma \in [0,1]$; however, it starts deviating as we increase $\gamma$ beyond 1. In contrast, the proposed framework, leveraging data-model fusion, sustains its accuracy even in regions extending three times beyond the training window. This clearly demonstrates the advantage of data-physics fusion over purely data-driven model.

\subsubsection{Arruda-Boyce}\label{sec:Ab}

\begin{figure}[hb!]
    \centering
    \begin{subfigure}{0.5\textwidth}
        \centering
        \includegraphics[width=\textwidth]{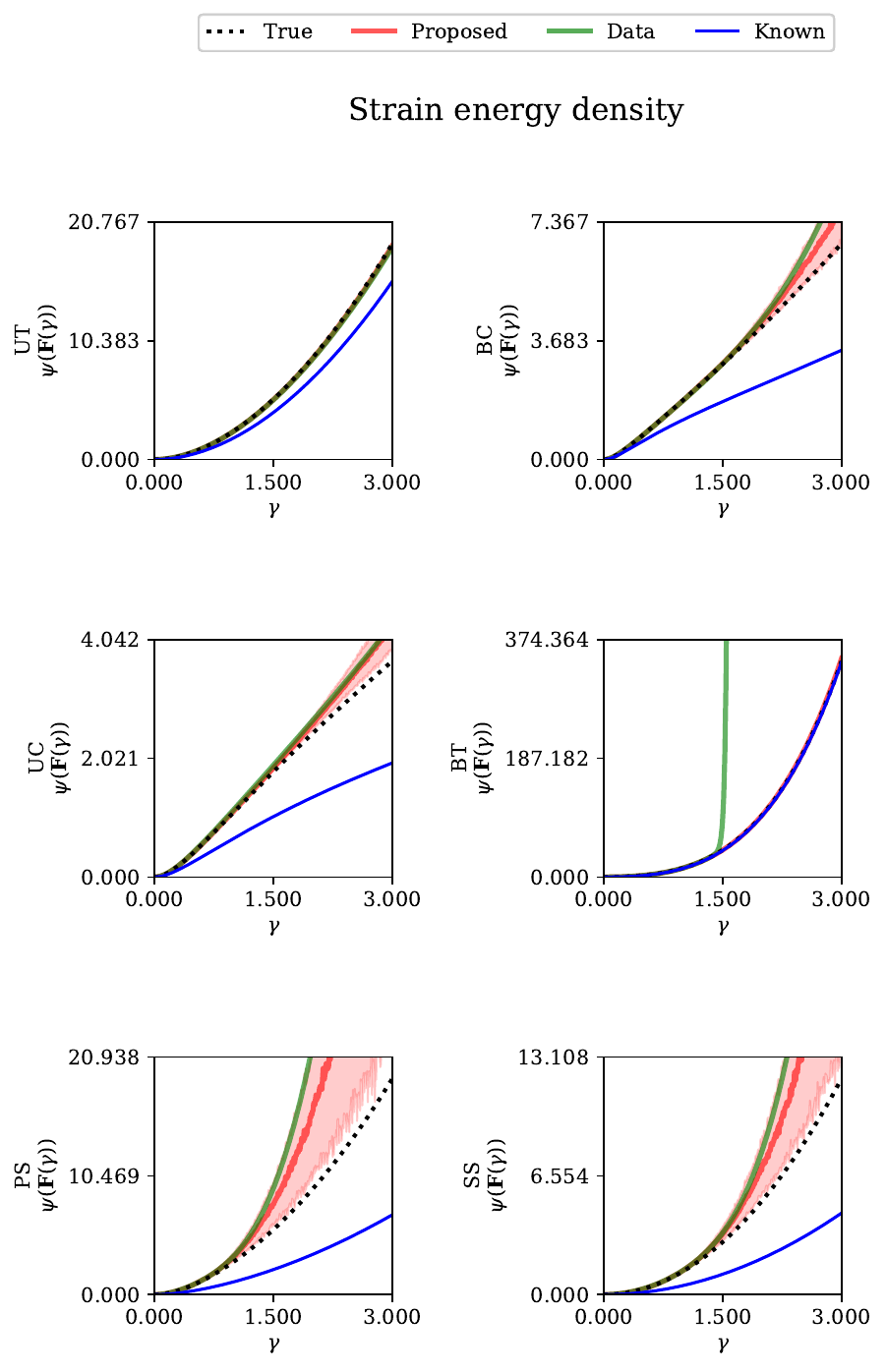}
        \caption{}
        \label{fig:se_n_a}
    \end{subfigure}%
    \begin{subfigure}{0.5\textwidth}
        \centering
        \includegraphics[width=\textwidth]{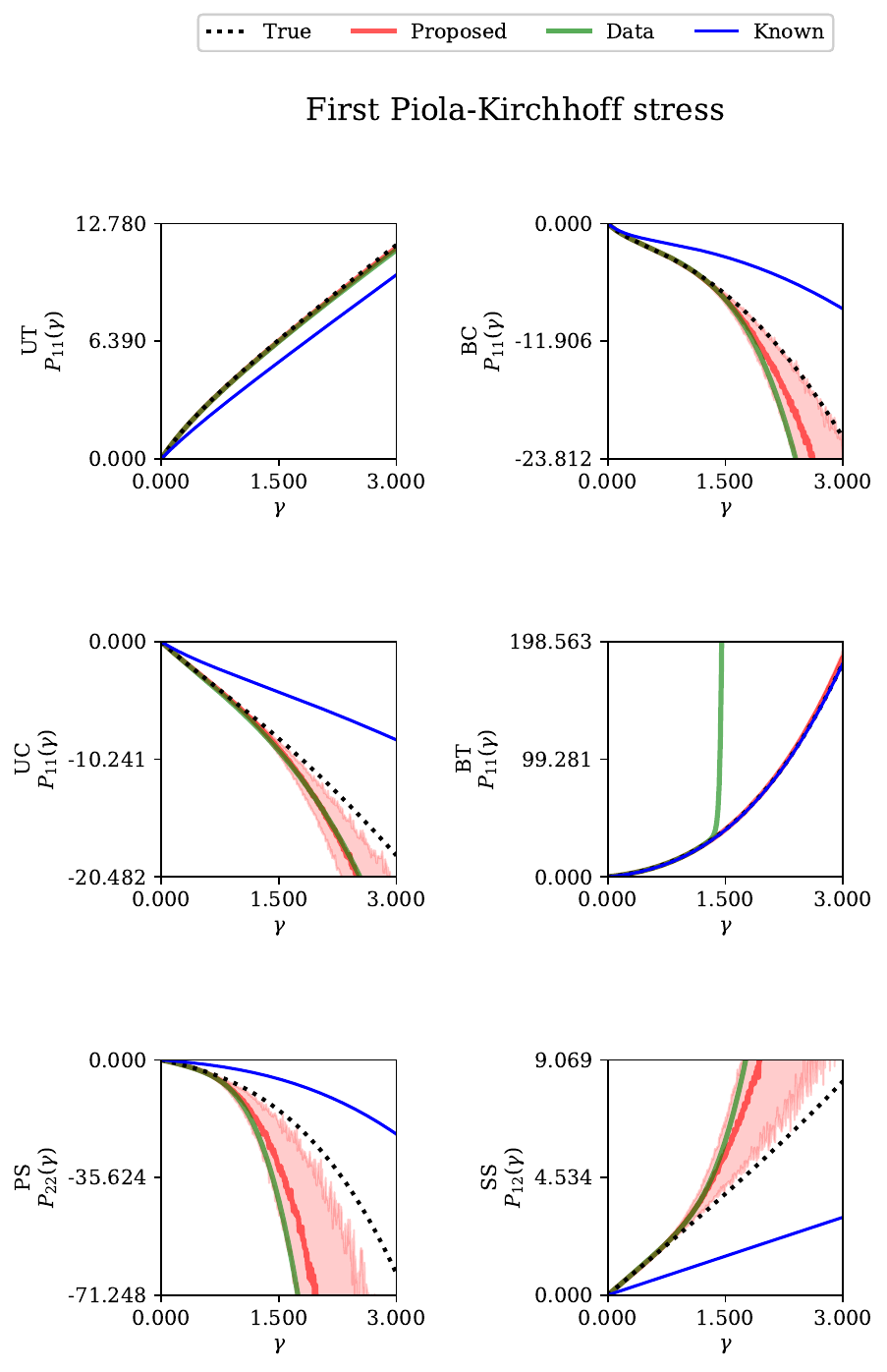}
        \caption{}
        \label{fig:pks_n_a}
    \end{subfigure}
    \caption{Comparison of predicted outputs: (a) Strain energy density \(\psi(F(\gamma))\) and (b) First Piola–Kirchhoff stress \(\bm{P}(\bm{F}(\gamma))\), produced by the proposed constitutive model, against a data-driven model without fusion (NN-Euclid) and a known phenomenological model, relative to the ground truth across six distinct stress states. The Neo-Hookean model is utilized as the known phenomenological model, along with the known fibre direction \(\alpha\), while the Arruda-Boyce model serves as the ground truth.}
    \label{fig:n_a}
\end{figure}

In this case, we have selected the Arruda-Boyce hyperelastic model as the ground truth. Developed by Arruda and Boyce \cite{arruda1993three}, this model is also referred to as the 8-chain model due to its distinctive network structure, which comprises eight chains extending from the centre of a cubic unit cell to its corners. The Neo-Hookean model (Eq.~\eqref{eq:neo}) is again employed as the known phenomenological model ($\psi_{\text{known}}$), and a deep learning model $\psi_{NN}$ is trained on a generalized state of stress, utilising data from only first six load steps, to learn the discrepancies. Again, we note that the model is trained based on displacement data by exploiting the differentiable solver that allows backpropagation through the solver.
To assess the model's extrapolation capabilities, we extended the loading parameter to \(\gamma \in [0, 3]\) and evaluated the trained model across six different stress states as before.

Figures \ref{fig:se_n_a} and \ref{fig:pks_n_a} show the strain energy density and Piola-Kirchhoff stress tensor as functions of the loading parameter \(\gamma\) across six deformation paths, comparing the proposed framework's performance with other models. The framework outperforms the other models, particularly in the extrapolated region where $\gamma > 1$. For the deformation paths of Pure Shear (PS) and Simple Shear (SS), where both the data-driven model and the known Neo-Hookean model deviate significantly from the ground truth in the extrapolative region, the proposed model closely aligns with the true values, nearly encompassing them. Notably, in the case of the uni-axial tension (UT), the proposed model follows its data-driven counterpart, which coincides with the ground truth. Similarly, for the bi-axial tension (BT), it aligns with its known phenomenological counterpart. This behaviour highlights the effectiveness of data-model fusion, leveraging the strengths of both approaches. It is also noteworthy that for the UT and BT deformation paths, the confidence intervals are narrow and are not discernible in the figures. This can be attributed to small variation in outputs for these deformation paths across all dropout masks, resulting in very small confidence intervals.

\subsubsection{Anisotropic solid with one fibre family oriented at $45^o$}\label{sec:An45}

\begin{figure}[h!]
    \centering
    \begin{subfigure}{0.5\textwidth}
        \centering
        \includegraphics[width=\textwidth]{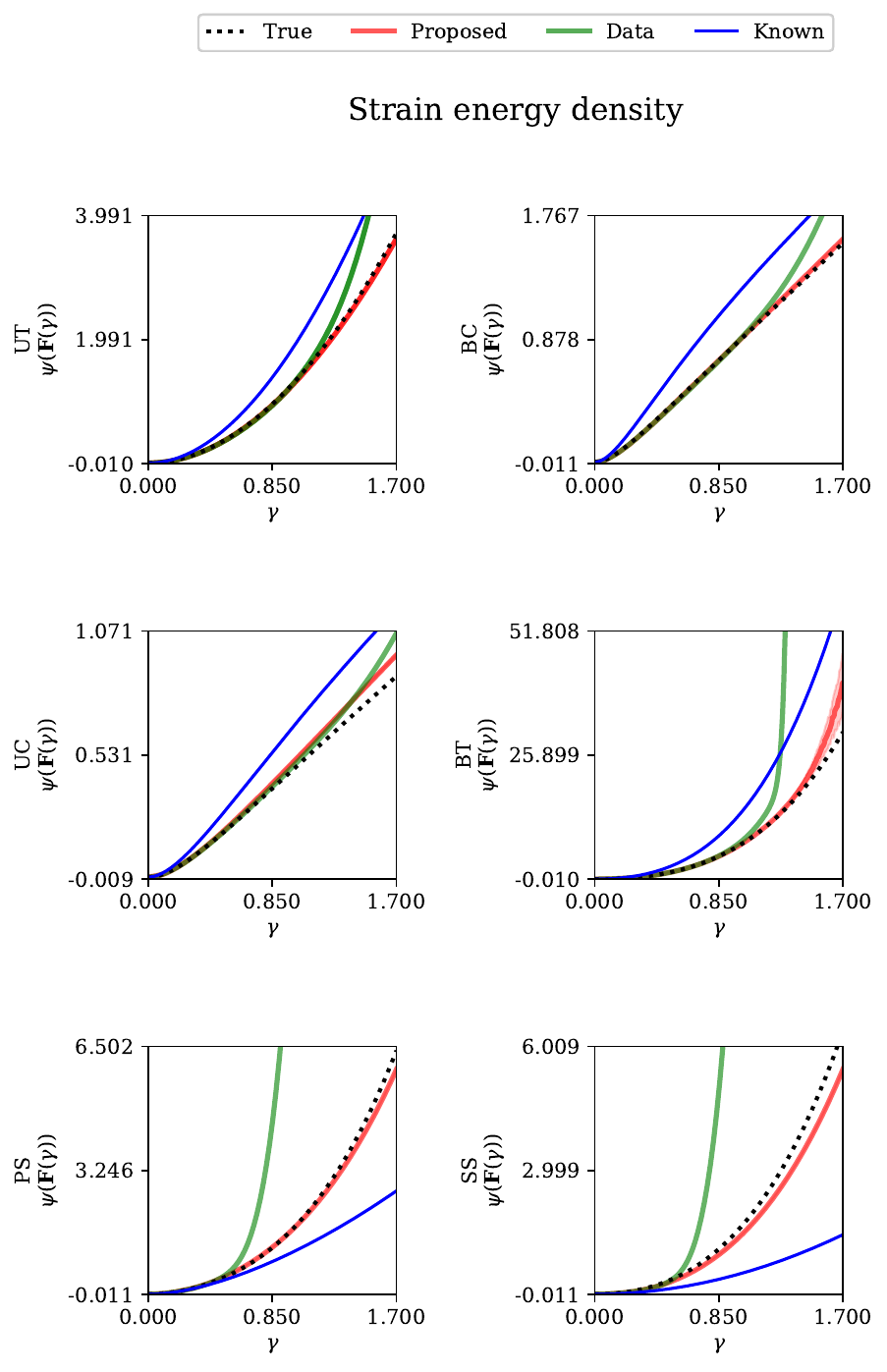}
        \caption{}
        \label{fig:se_n_a45_gvn_a}
    \end{subfigure}%
    \begin{subfigure}{0.5\textwidth}
        \centering
        \includegraphics[width=\textwidth]{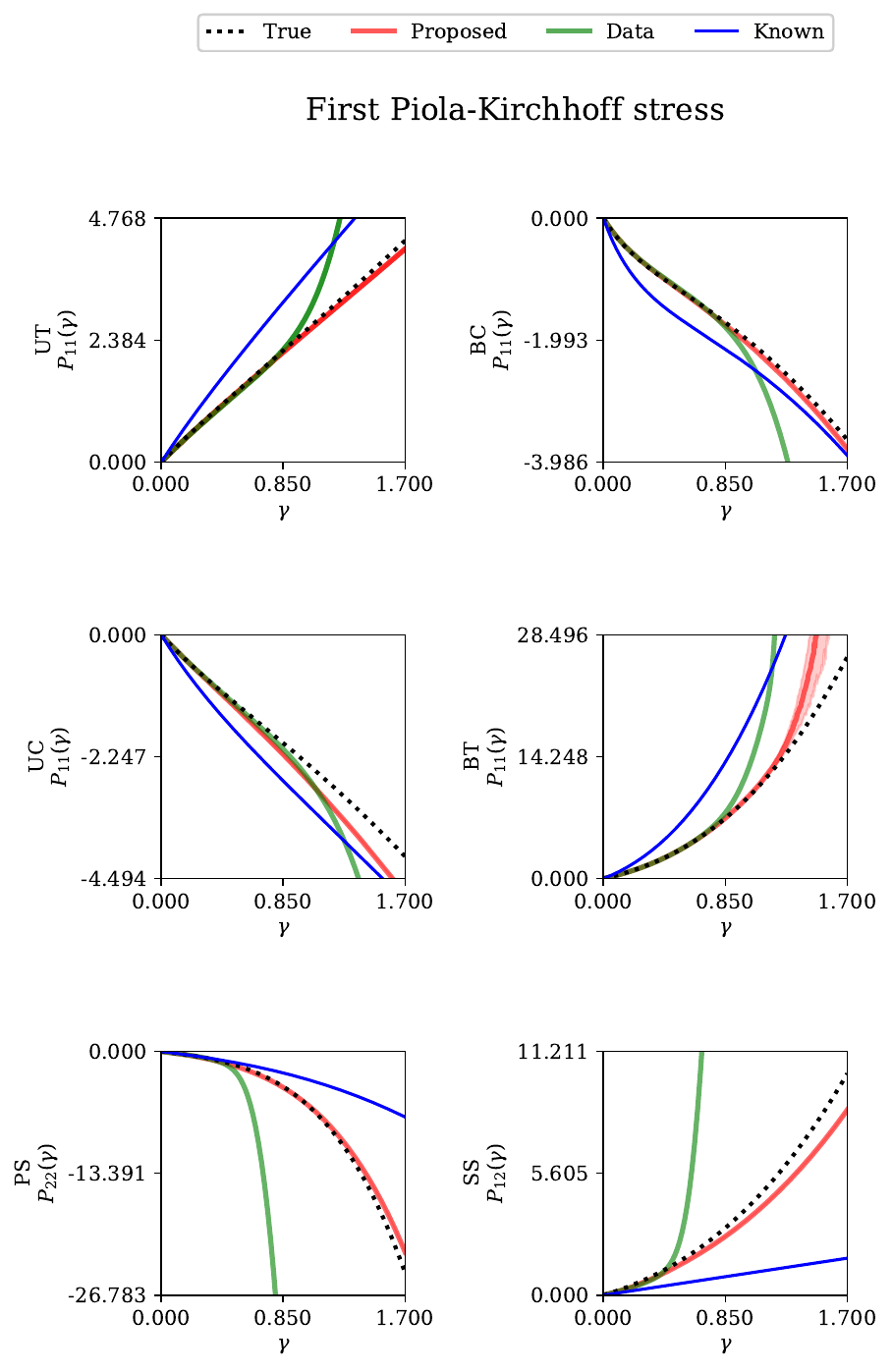}
        \caption{}
        \label{fig:pks_n_a45_gvn_b}
    \end{subfigure}
    \caption{Comparison of predicted outputs: (a) Strain energy density \(\psi(F(\gamma))\) and (b) First Piola–Kirchhoff stress \(\bm{P}(\bm{F}(\gamma))\), produced by the proposed constitutive model, against a data-driven model without fusion (NN-Euclid) and a known phenomenological model, relative to the ground truth across six distinct stress states. The Neo-Hookean model is utilized as the known phenomenological model, along with the known fibre direction \(\alpha\), while the anisotropic model with a single fibre family aligned at \(\alpha = 45^\circ\) serves as the ground truth.
}
    \label{fig:n_a45_gvn_a}
\end{figure}

\begin{figure}[h!]
    \centering
    \begin{subfigure}{0.5\textwidth}
        \centering
        \includegraphics[width=\textwidth]{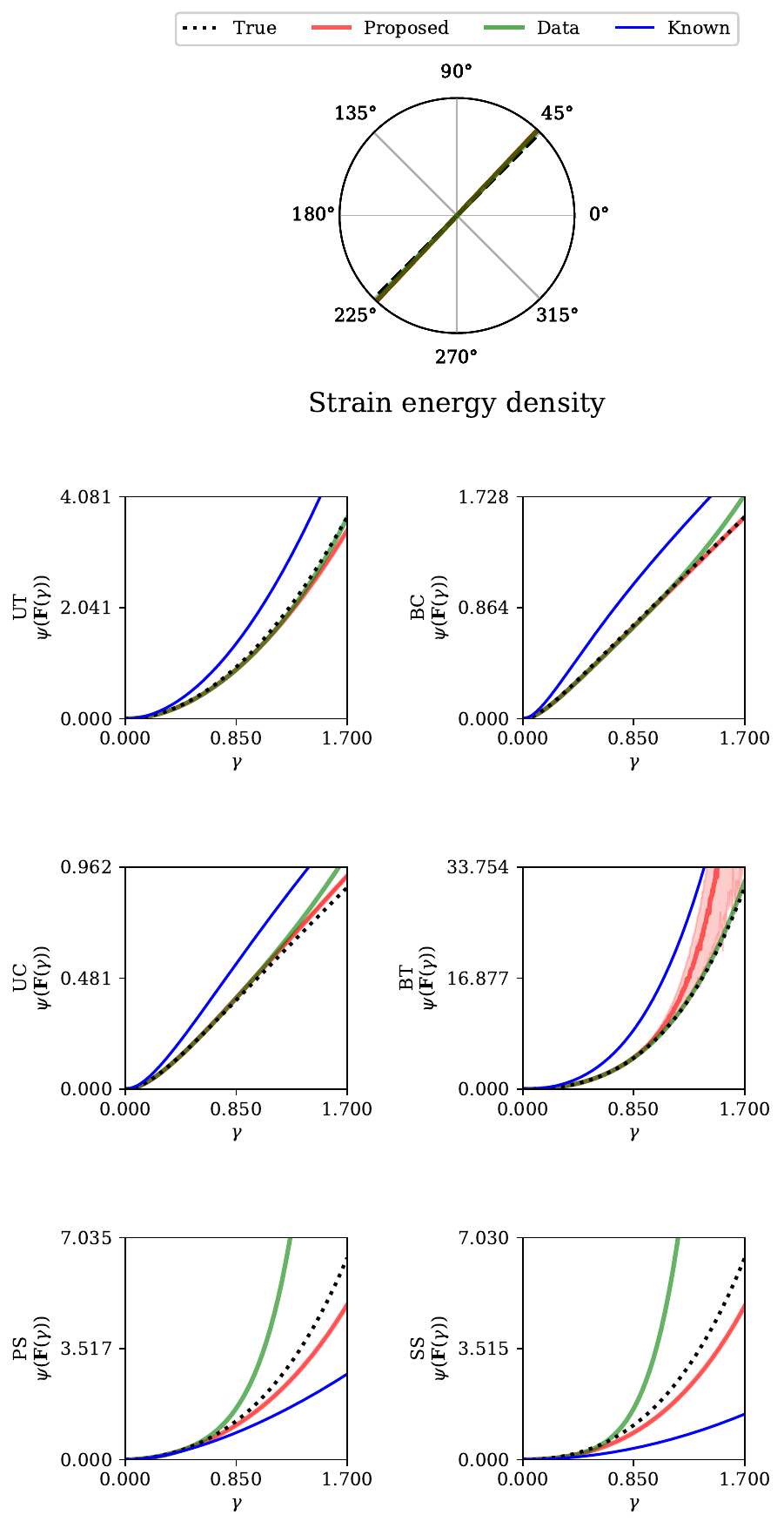}
        \caption{}
        \label{fig:se_n_a45}
    \end{subfigure}%
    \begin{subfigure}{0.5\textwidth}
        \centering
        \includegraphics[width=\textwidth]{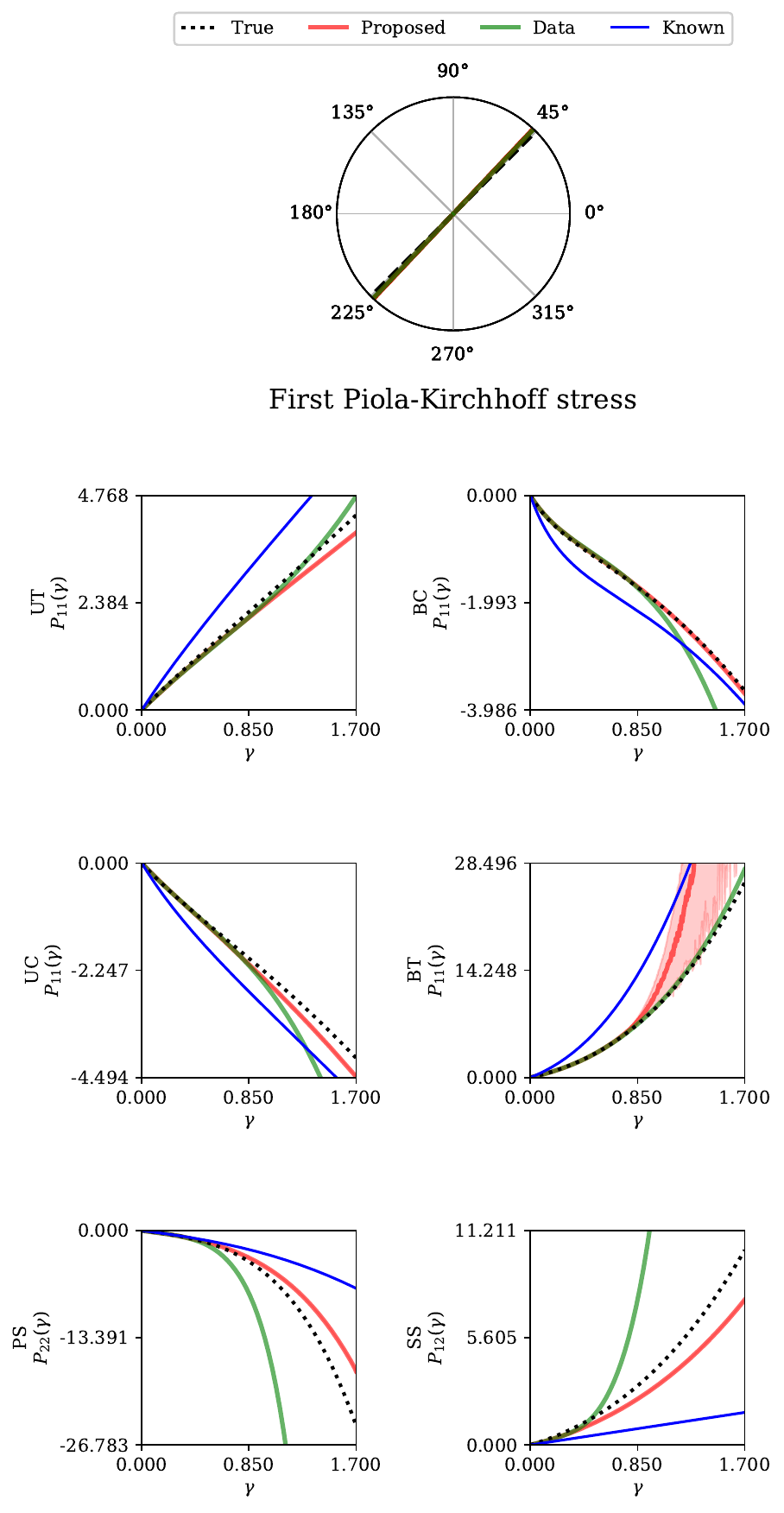}
        \caption{}
        \label{fig:pks_n_a45}
    \end{subfigure}
    \caption{Comparison of predicted outputs: (a) Strain energy density \(\psi(F(\gamma))\) and (b) First Piola–Kirchhoff stress \(\bm{P}(\bm{F}(\gamma))\), produced by the proposed constitutive model, against a data-driven model without fusion (NN-Euclid) and a known phenomenological model, relative to the ground truth across six distinct stress states. The Neo-Hookean model is utilized as the known phenomenological model while the anisotropic model with a single fibre family aligned at \(\alpha = 45^\circ\) serves as the ground truth.}
    \label{fig:n_a45}
\end{figure}
In this case, we selected an anisotropic material with a single fibre family oriented at 45 degrees as our ground truth. The Neo-Hookean model (Eq.~\eqref{eq:neo}) is again used as the known phenomenological model, $\psi_{\text{known}}$, while a deep learning model, $\psi_{NN}$, is trained on generalized stress states using data from only the first five load steps to capture the discrepancies. A key aspect of this anisotropic case is the fibre orientation, denoted as \(\alpha\).
Two scenarios were investigated: in the first, the fibre direction \(\alpha\) was predefined; in the second, 
it was treated as a learnable parameter, as described below:
\[
\psi = \psi(\bm{F}; \bm{\theta}_\alpha) = \psi_{\text{known}}(\bm{F}) + \psi_{\text{NN}}(\bm{F}; \bm{\theta}_\alpha) + \psi_{\text{energy}} + \psi_{\text{stress}}
\]
where \(\bm{\theta}_\alpha\) represents the set of learnable parameters, including \(\alpha\).
It is introduced as an input to the neural network with the invariant $(\tilde{I}_{\alpha} - 1)^2$ where $\tilde{I}_{\alpha} =  J^{-2/3} (\mathbf{a} \cdot \bm{C} \mathbf{a})$ and $\mathbf{a} =(\cos \alpha, \sin \alpha, 0)^T$ as mentioned earlier. The network is trained to optimize both \(\alpha\) and the weights and biases \(\theta\) using gradient-based methods and gradients with respect to both \(\alpha\) (\(\frac{\partial \mathcal{L}}{\partial \alpha}\)) and \(\bm \theta\)( \(\frac{\partial \mathcal{L}}{\partial \bm \theta}\)) are computed to minimize the loss function during training. Furthermore, the loading parameter \(\gamma\) was extended beyond its training range up to \(\gamma \in [0, 1.7]\) in both scenarios to assess the framework's extrapolation capabilities.

In the first scenario, Figures \ref{fig:se_n_a45_gvn_a} and \ref{fig:pks_n_a45_gvn_b} illustrate the good alignment of the proposed approach with the ground truth, clearly outperforming other models across all stress states, particularly in the extrapolated regions.

In the second scenario, Figures \ref{fig:se_n_a45} and \ref{fig:pks_n_a45} highlight the performance of the proposed framework across all the stress states, demonstrating a close alignment with the ground truth in both strain energy density and Piola-Kirchhoff stress plots. In this case, accurate learning of fibre direction \(\alpha\) is crucial for the accuracy of the model; therefore, we incorporated polar plots alongside the \(\psi(\gamma)\) and \(P(\gamma)\) plots to compare the learned fibre direction \(\alpha\) from the data-driven and proposed approaches relative to the ground truth. It is noteworthy that the Neo-Hookean phenomenological model, lacking directional information, does not contribute to the polar plot. The results indicate that the learned \(\alpha\) closely matches the true \(\alpha\), as evidenced in the polar plot and the proposed framework significantly outperforms the other models in the \(\psi(\gamma)\) and \(P(\gamma)\) plots,  particularly in the extrapolated region where \(\gamma > 1\). Additionally, the confidence interval in most stress states is too narrow to be noticeable visually. This is likely due to very small variation in the prediction outputs for these stress states across the dropout masks, resulting in very small confidence intervals.
\subsection{FEM Analysis}

\begin{figure}[ht!]
\captionsetup[subfigure]{labelformat=empty}
  \centering
  \begin{tabular}{ccc}
  \begin{subfigure}{0.2\textwidth}
      \includegraphics[height=3.3 cm]{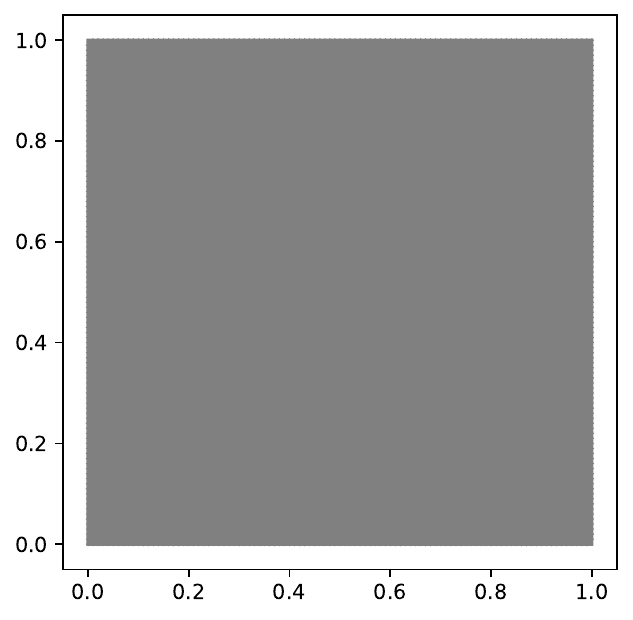}
      \caption{Mesh-1}
    \end{subfigure} &
    \begin{subfigure}{0.2\textwidth}
      \includegraphics[width=\linewidth]{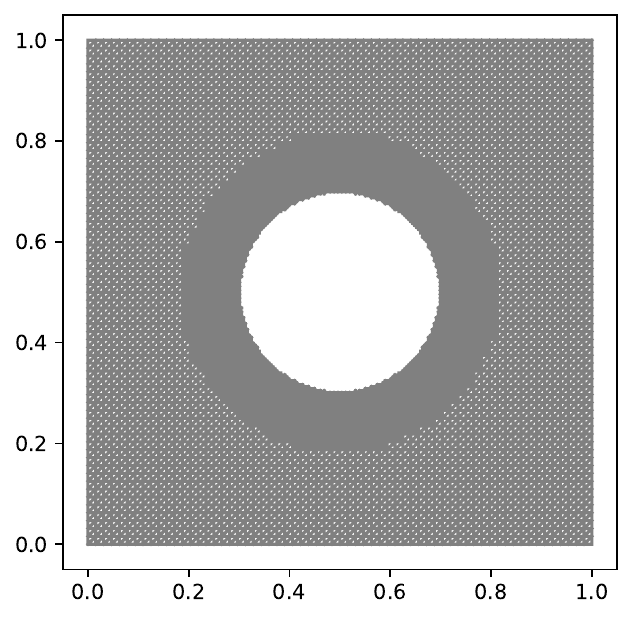}
      \caption{Mesh-2}
    \end{subfigure} &
    \begin{subfigure}{0.2\textwidth}
      \includegraphics[width=\linewidth]{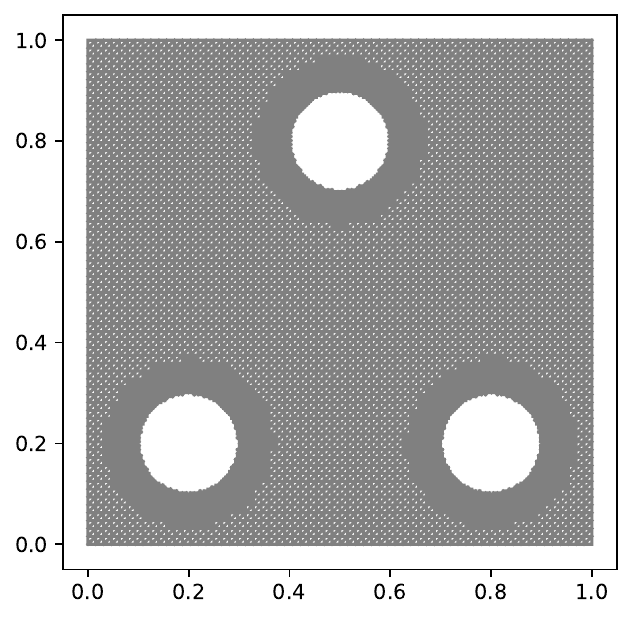}
      \caption{Mesh-3}
    \end{subfigure} \\
    \begin{subfigure}{0.2\textwidth}
      \includegraphics[width=\linewidth]
      {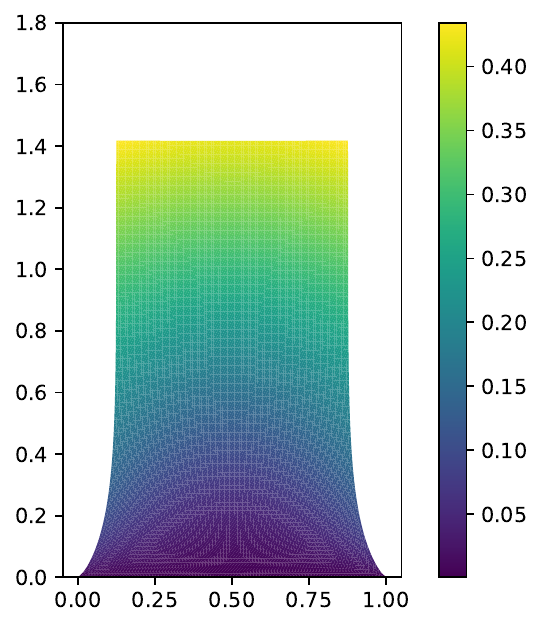}
      \caption{Ground Truth-1}
    \end{subfigure} &
    \begin{subfigure}{0.2\textwidth}
      \includegraphics[width=\linewidth]{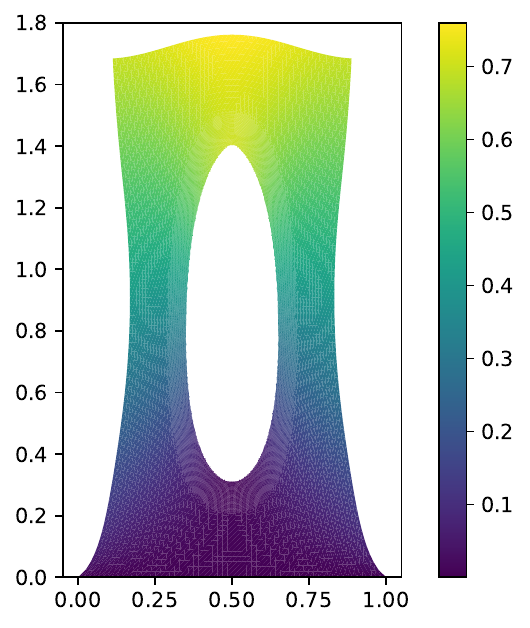}
      \caption{Ground Truth-2}
    \end{subfigure} &
    \begin{subfigure}{0.2\textwidth}
      \includegraphics[width=\linewidth]{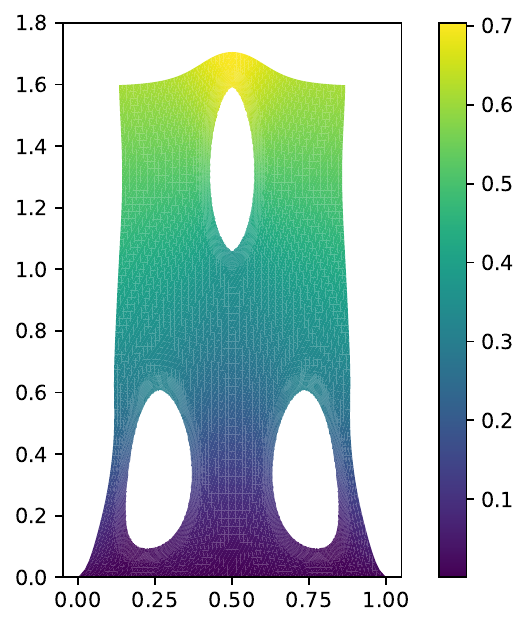}
      \caption{Ground Truth-3}
    \end{subfigure} \\
    \begin{subfigure}{0.2\textwidth}
      \includegraphics[width=\linewidth] {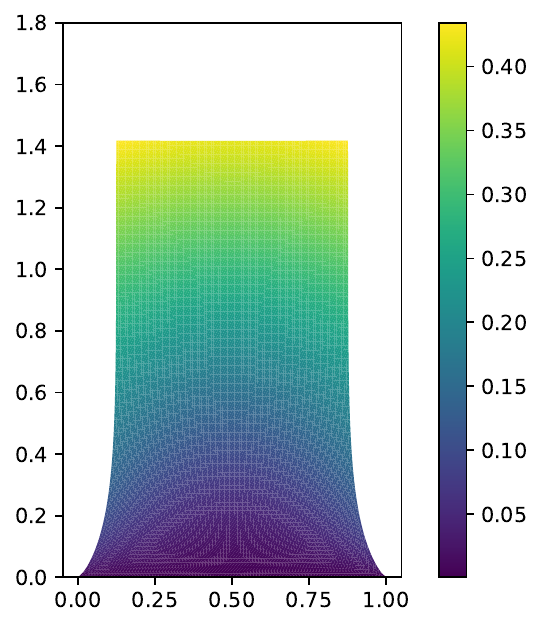}
      \caption{Proposed-1}
    \end{subfigure} &
    \begin{subfigure}{0.2\textwidth}
      \includegraphics[width=\linewidth]{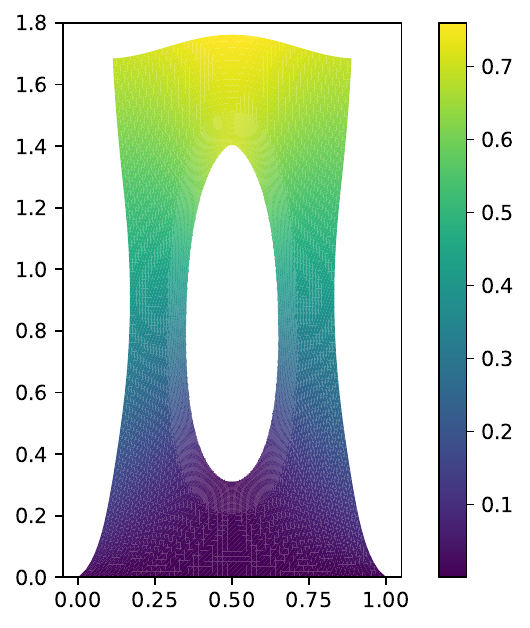}
      \caption{Proposed-2}
    \end{subfigure} &
    \begin{subfigure}{0.2\textwidth}
      \includegraphics[width=\linewidth]{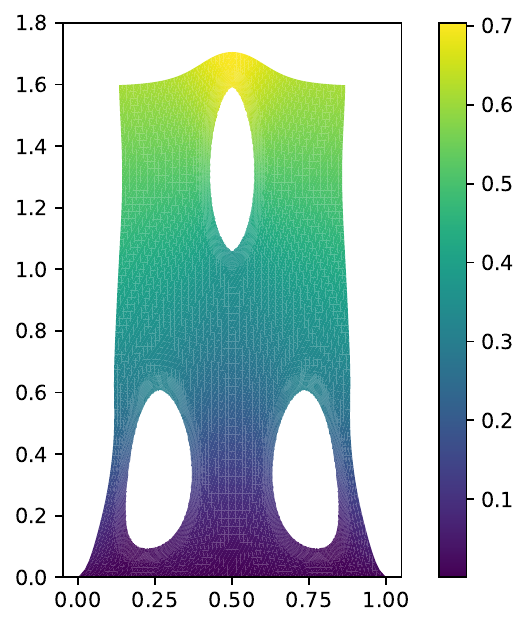}
      \caption{Proposed-3}
    \end{subfigure}\\
  \end{tabular}
  \caption{FEM simulation results, using the proposed constitutive model with Neo-Hookean as the known phenomenological model and Isihara model as the ground truth.}
  \label{fig:fem}
\end{figure}
We also performed finite element method (FEM) analysis on three geometries of increasing complexity, to demonstrate the practical applicability of the proposed approach. First, we considered a simple square plate with dimensions \(1.0 \, \text{m} \times 1.0 \, \text{m}\). In the second case, we introduced a single circular hole of radius \(0.2 \, \text{m}\) at the centre of the square plate, maintaining all other conditions and dimensions the same. This modification increased the complexity of the analysis by introducing stress concentration around the hole. Lastly, to further increase the complexity, we introduced three smaller holes of radius \(0.1 \, \text{m}\) to the square plate of the same dimension. Two holes are at the bottom (bottom-left and bottom-right) and one is at the top centre. The centers of the bottom holes were positioned at \((0.2 \, \text{m}, 0.2 \, \text{m})\) and \((0.8 \, \text{m}, 0.2 \, \text{m})\), while the top hole was centered at \((0.5 \, \text{m}, 0.8 \, \text{m})\). In all three cases, the bottom edge of the plate was fixed (clamped condition i.e. (\(u = 0\)). The top edge of the plate was subjected to an external pressure load, gradually applied in small increments up to a maximum pressure of \(1.0 \, \text{Pa}\). This step-wise loading approach ensured better convergence in the non-linear deformation of the plates.

The domain for each geometry was discretized using a mesh with 64 elements in each direction. For the cases with holes, additional mesh refinement was performed around the holes to better capture the stress concentrations. The FEM analysis was performed using the FEniCS package, incorporating the proposed framework as the constitutive model. The results were benchmarked against those derived from the exact Isihara phenomenological model, as illustrated in Fig. \ref{fig:fem}. The results indicate that the FEM analysis using the proposed framework matches well with the phenomenological model, demonstrating the effectiveness of the proposed approach in handling complex geometries and loading conditions.

\section{Conclusion}\label{sec:concl}
In this paper, we present a novel hybrid framework, "Fusion-based Constitutive model (FuCe)",
designed for constitutive modelling with a focus on hyperelastic materials. Our approach seamlessly integrates existing phenomenological constitutive models with Input Convex Neural Networks (ICNNs), enabling the deep learning model to capture the discrepancy between the traditional phenomenological model and the actual material behaviour as observed in the data. This method not only corrects the phenomenological model based on data but also preserves the model's interpretability and generalization capabilities.

We applied the proposed framework to two isotropic and one anisotropic materials material. The framework was rigorously evaluated under six different stress states and further validated through finite element method (FEM) simulations for Isihara material to assess its practical applicability. The key findings from this study are summarized below:

\begin{itemize}
    \item The proposed framework demonstrates good accuracy in all the example problems, along with the FEM simulation. The accuracy is maintained even for extrapolations in terms of loading parameters. This showcases the robustness, and generalisation capacity of the proposed framework.
    \item The proposed framework satisfies all essential constitutive requirements, including material objectivity, stress symmetry, local polyconvexity, and normalization conditions. Some of these are inherently achieved by construction, while others are ensured by incorporating physics-based corrective terms. This approach guarantees stress-free reference states and aligns the model closely with the behaviour of real-world materials
    \item The proposed model is trained using practically feasible data, consisting of sparse and noisy full-field displacement measurements and global reaction force data. This training process leverages a differentiable solver, enabling backpropagation through the solver. In the absence of stress labels, the model's learning is guided by a physics-informed loss function based on the conservation of linear momentum, effectively compensating for the lack of direct stress measurements.
    \item The proposed framework also quantifies predictive uncertainty by incorporating Monte Carlo dropout, a simple yet effective approximate Bayesian method. This approach enhances the dependability of the model's predictions and supports more informed decision-making.
\end{itemize}

In summary, the proposed FuCe framework successfully integrates classical phenomenological models with emerging deep learning approaches for constitutive modelling of complex materials, such as hyperelastic materials, as investigated in this study. It demonstrates strong alignment with ground truth data, even in extrapolative regions, and is satisfactorily applicable within FEM simulations, producing accurate results. However, some deviations were observed, particularly for the biaxial tensile stress-state. One potential solution is to fine-tune FuCe for biaxial data. Additionally, we have only focused on hyperelastic material model; other constitutive relations have not been investigated in this work. 
Future work will focus on addressing these challenges.

\section*{Acknowledgements}
 SC acknowledges the financial support received from Anusandhan National Research Foundation (ANRF) via grant no. CRG/2023/007667. SC and RN acknowledges the financial support received from the Ministry of Port and Shipping via letter no. ST-14011/74/MT (356529). 

\section*{Code availability}
Upon acceptance, all the source codes to reproduce the results in this study will be made available on request

\section*{Competing interests} 
The authors declare no competing interests.
\clearpage


\end{document}